\documentclass[useAMS,usenatbib]{mn2e}
\usepackage{psfig, epsf, epsfig}

\title[The Galactic microquasar GRS 1758$-$258]{Accretion states 
of the Galactic microquasar GRS~1758$-$258}
\author[R. Soria et al.]{Roberto Soria$^{1,2}$\thanks{E-mail:
roberto.soria@mssl.ucl.ac.uk},
Jess W. Broderick$^{3}$, 
JingFang Hao$^{4}$, 
Diana C. Hannikainen$^{5}$, 
\newauthor
Missagh Mehdipour$^{1}$,
Katja Pottschmidt$^{6,7}$
and Shuang-Nan Zhang$^{8}$\\
$^{1}$Mullard Space Science Laboratory, University College London, Holmbury St Mary, Surrey RH5 6NT, UK\\
$^{2}$Curtin Institute of Radio Astronomy, Curtin University, GPO Box U1987, Perth, WA 6845, Australia\\
$^{3}$School of Physics \& Astronomy, University of Southampton, Southampton, Hampshire SO17 1BJ, UK\\
$^{4}$Department of Physics and Tsinghua Center for Astrophysics, Tsinghua University, 
Beijing 100084, China \\
$^{5}$Aalto University Mets\"ahovi Radio Observatory,  Mets\"ahovintie 114, FI-02540 Kylm\"al\"a, Finland\\
$^{6}$CRESST and NASA Goddard Space Flight Center, Astrophysics Science Division, Code 661, Greenbelt, MD 20771, USA\\
$^{7}$Center for Space Science and Technology, University of Maryland, 1000 Hilltop Circle, Baltimore, MD 21250, USA\\
$^{8}$Key Laboratory of Particle Astrophysics, Institute of High Energy Physics, 
Chinese Academy of Sciences, Beijing 100049, China}

\begin{document}

\date{Accepted 2011 March 15 --- Received 2011 March 05 --- in original form 2011 February 1}

\pagerange{\pageref{firstpage}--\pageref{lastpage}} \pubyear{2010}

\maketitle

\label{firstpage}

\begin{abstract}
We present the results of a radio and X-ray study of the Galactic microquasar 
GRS 1758$-$258, using unpublished archival data and new observations. 
We focus in particular on the 2000--2002 state transitions, and on its 
more quiet behaviour in 2008--2009. Our spectral and timing analysis 
of the {\it XMM-Newton} data shows that the source was in the canonical 
intermediate, soft and hard states in 2000 September 19, 2001 March 22 and 2002 September 28,
respectively. We estimate the disk size, luminosity and temperature, which are 
consistent with a black hole mass $\sim 10 M_{\odot}$. There is much overlap 
between the range of total X-ray luminosities (on average $\sim 0.02 L_{\rm Edd}$) 
in the hard and soft states, and probably between the corresponding 
mass accretion rates; in fact, the hard state is often more luminous.
The extended radio lobes seen in 1992 and 1997 are still present in 2008--2009.
The 5-GHz radio core flux density has shown variability between 
$\sim 0.1$--$0.5$ mJy over the last two decades. This firmly places 
GRS 1758$-$258 in the radio-quiet sequence of Galactic black holes, 
in the radio/X-ray plane. We note that this dichotomy is similar 
to the dichotomy between the radio/X-ray sequences of Seyfert  
and radio galaxies. We propose that the different radio efficiency 
of the two sequences is due to relativistic electron/positron jets 
in radio-loud black holes, and sub-relativistic, thermally dominated 
outflows in radio-quiet sources. 
\end{abstract}

\begin{keywords}
accretion, accretion discs --- X-rays: binaries --- X-rays: individual: GRS 1758$-$258  --- 
radio continuum: general --- black hole physics.
\end{keywords}

\section{Introduction}

Understanding the connection between black hole (BH) accretion  
and jet properties is a fundamental astrophysical problem.
It is the key for the modelling of BH growth and feedback effects.
Much progress has been achieved over the last two decades, 
thanks to combined X-ray/radio studies of transient 
stellar-mass BHs, over timescales of a few months/years. 
Taken together, X-ray spectral and timing properties, and 
radio jet properties are the empirical criteria 
that identify physical ``accretion states'' \citep{fen04,mcc06}.

When the mass accretion rate is below a few percent of 
the classical Eddington rate, the accretion flow is hotter, 
geometrically thick, can more easily support a poloidal magnetic 
field and accelerate/collimate outflows. Radiative efficiency 
decreases and mechanical efficiency (jet power) increases.
Advection-dominated flows may set in (ADAF models: \citealt{nar94,esi97}), 
due to inefficient cooling at low densities.
Observationally, the hard state is characterized 
by flat-spectrum radio emission from the core (direct signature 
of a continuous, self-absorbed jet), by an X-ray spectrum dominated 
by a hard power-law (photon index $\Gamma \sim 1.7$), and by high 
short-term variability in the power density spectrum.

At accretion rates above a few percent of the Eddington rate, 
most stellar-mass BHs are seen to switch to the thermal-dominant 
(high/soft) state, characterised by an optically thick, 
geometrically thin disk. This state is radiatively efficient 
and the jet is suppressed: these BHs are radio quiet. The X-ray 
emission comes mostly from the inner part of the disk, 
near the innermost stable circular orbit. The inner radius 
and peak temperature of the disk depend directly on the accretion 
rate and BH mass. 

However, the physical mechanisms that cause BHs to switch between 
the hard and soft state are still poorly understood, 
as are the intermediate states between them. The sample of transient 
BHs (mostly Galactic BHs with a low-mass donor star), used to define 
the ``canonical'' state behaviour may be biased.
The luminosity threshold for the collapse of the thick flow 
and the suppression of the jet changes not only between different sources, 
but also for individual sources between different outbursts, 
and between the rising and declining phase of the state transition 
cycle, due to hysteresis---once it is formed, it takes time for 
the accretion disk to dissipate after the accretion rate 
has dropped. Furthermore, some BHs (particularly among 
the ultraluminous X-ray source class: \citealt{sor11}) do not appear to switch 
to the thermal-dominant state even at high luminosities, and 
have an X-ray spectrum persistently dominated by a power-law-like 
component. The origin of the power law component itself is still unclear, 
with possible contributions from thermal Comptonization 
in a hot corona, and synchrotron self-Compton emission 
from the base of the jet \citep{mar05}.

In this paper, we investigate accretion states  
by studying the X-ray and radio properties 
of the Galactic microquasar GRS 1758$-$258 
\citep{man90,sun91}, which stands out among Galactic BHs 
for its persistent activity, mostly in a moderately luminous hard state. 
We focus on its behaviour during 2000--2002 (Figure 1), when it underwent a series 
of spectral state transitions, and in 2008--2009 (Figure 2), when it was in a more 
steady hard state. For the X-ray study, we use unpublished {\it XMM-Newton}  
data supplemented by {\it Swift} data. For the radio study, we use 
our recent observation from the Australia Telescope Compact Array (ATCA), 
supplemented by archival Very Large Array (VLA) data.



\begin{figure}
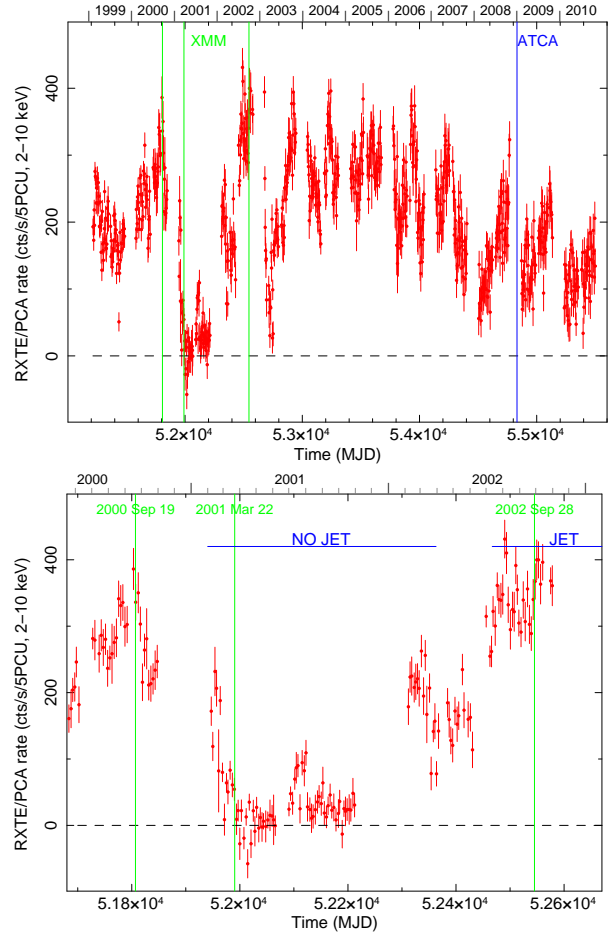

\begin{center}
\psfig{figure=pca_lc_paper_rev.ps,width=62mm,angle=270}
\psfig{figure=pca_lc_paper2_rev.ps,width=62mm,angle=270}
\end{center}
\caption{Top panel: {\it RXTE}/PCA lightcurve of GRS 1758$-$258 
between 1999 and 2010, from the Galactic Bulge Scans, and timeline 
of our {\it XMM-Newton} and ATCA observations. Bottom panel: zoomed-in 
view around the 2000--2002 {\it XMM-Newton} observations, which coincided 
with a transition to the soft state and back to the hard state. 
We have also marked the epochs when a compact jet was or was not detected 
in the VLA data. Galactic Bulge Scan lightcurves are provided by Craig 
Markwardt and are available at http://lheawww.gsfc.nasa.gov/users/craigm/galscan/main.html~.}
\label{f1}
\end{figure}

\begin{figure}
\begin{center}
\psfig{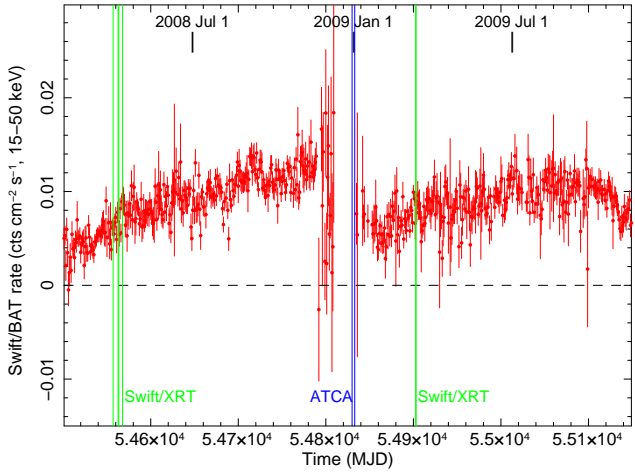}
\end{center}
\caption{{\it Swift}/BAT lightcurve during 2008--2009, and timeline 
of the {\it Swift}/XRT spectral observations and of our ATCA observations. 
{\it Swift}/BAT lightcurves are provided by Hans Krimm and are available at 
http://heasarc.nasa.gov/docs/swift/results/transients/.
The scarcity or poor quality of X-ray coverage during 
most of December and January is due to an unfavourable 
solar orientation.}
\label{f2}
\end{figure}

\section{Main properties of GRS 1758$-$258}


The BH candidate GRS 1758$-$258 is located in the direction 
of the Galactic centre, $\approx 5^{\circ}$ away from it on the sky. 
Here we shall assume a distance of $8$ kpc.
Its BH mass and orbital period are still unknown, because optical 
studies of the stellar donor are hampered by high extinction 
\citep{rot02,smi02b,mun10}.
The optical counterpart has a dereddened luminosity consistent 
with an early A-type main sequence star, but unusual colours \citep{mun10}; 
part of the optical/IR emission may also come from the accretion disk or the jet. 

GRS J1758$-$258 has a point-like core with radio and X-ray emission, 
and an extended lobe structure, with a projected distance of 
$\approx 2.5$ arcmin between the lobes ($\approx 6$ pc at $d=8$ kpc).
The radio core is variable or transient. For example, we already knew that 
it was undetected ($\la 0.1$ mJy) at 6 cm in early 1992 \citep{rod92}, 
when the BH was in an X-ray soft state. It was detected at variable flux 
levels $\approx 0.1$--$0.5$ mJy in 1992 September, when the BH switched 
back to the hard state, in 1993 and in 1997 \citep{mir94,mar02,har05}. 
The extended radio lobes, with an optically-thin synchrotron spectrum, 
appear to be persistent structures rather than fast-moving 
ejections \citep{har05}. This is slightly unusual 
for Galactic microquasars \citep{heinz02}. 

In contrast with a sparse radio coverage, 
there has been an almost continuous hard X-ray monitoring over the last two decades 
with {\it GRANAT}, {\it CGRO}, {\it RXTE}, {\it INTEGRAL} and {\it Swift} 
\citep{lin00,smi02a,smi02b,pot06,loh11}. There was also   
a series of four {\it Chandra} observations in 2000--2001 \citep{hei02a}, 
and three {\it XMM-Newton} observations in 2000--2002 \citep{gol01,mil02}.
The X-ray data show that the source is always active (unusual 
for Galactic BHs with low-mass donor stars), with a variability 
by a factor of a few. It has spent most of the last two decades 
in a hard state, with occasional transitions to an intermediate 
or soft state. The typical $0.3$--$10$ keV luminosity 
is $\approx 1$--$2 \times 10^{37}$ erg s$^{-1}$, corresponding 
to accretion rates $\sim 0.01$--$0.03$ Eddington 
for typical stellar-mass BHs with masses $\sim 10 M_{\odot}$ 
and radiative efficiency $\sim 0.1$.




\begin{table}
\begin{center}
\begin{tabular}{lcccr}
\hline
Date & $F_{\rm 4.9GHz}$ & RMS Noise & Beam  & Beam  \\
     & ($\mu$Jy) & ($\mu$Jy/beam) &  FWHM & PA\\
\hline\\[-5pt]
2001-02-01  & $<240$ & $80$  & $7\farcs10 \times 0\farcs54$ & $49\fdg4$ \\
2001-02-06  & $<393$ & $131$  & $3\farcs45 \times 0\farcs64$ & $-53\fdg1$ \\
2001-03-14  & $<213$ & $71$  & $3\farcs30 \times 1\farcs12$ & $-28\fdg3$ \\
2001-03-22  & $<105$ & $35$  & $2\farcs35 \times 1\farcs13$ & $1\fdg6$ \\
2001-03-25  & $<132$ & $44$  & $2\farcs41 \times 1\farcs13$ & $-5\fdg3$ \\
2001-04-16  & $<156$ & $52$  & $2\farcs35 \times 1\farcs12$ & $2\fdg1$ \\
2001-04-29  & $<189$ & $63$  & $2\farcs48 \times 1\farcs12$ & $-11\fdg8$ \\
2001-05-19  & $<189$ & $63$  & $2\farcs47 \times 1\farcs15$ & $-7\fdg7$ \\
2001-06-05  & $<249$ & $83$  & $5\farcs44 \times 2\farcs45$ & $-59\fdg7$ \\
2001-06-11  & $<213$ & $71$  & $4\farcs10 \times 2\farcs60$ & $68\fdg8$ \\
2001-06-18  & $<216$ & $72$  & $4\farcs20 \times 2\farcs64$ & $68\fdg8$ \\
2001-07-08  & $<273$ & $91$  & $9\farcs98 \times 3\farcs63$ & $18\fdg1$ \\
2001-07-24  & $<273$ & $91$  & $9\farcs75 \times 3\farcs79$ & $16\fdg4$ \\
2001-08-04  & $<252$ & $84$  & $8\farcs35 \times 4\farcs13$ & $13\fdg5$ \\
2001-08-09  & $<246$ & $82$  & $8\farcs36 \times 3\farcs63$ & $-3\fdg9$ \\
2001-08-16  & $<264$ & $88$  & $8\farcs18 \times 3\farcs55$ & $2\fdg2$ \\
2001-08-26  & $<279$ & $93$  & $8\farcs91 \times 3\farcs63$ & $-12\fdg1$ \\
2001-08-30  & $<276$ & $92$  & $9\farcs26 \times 3\farcs69$ & $16\fdg9$ \\
2001-09-07  & $<255$ & $85$  & $9\farcs39 \times 3\farcs66$ & $19\fdg8$ \\
2001-11-05  & $<324$ & $108$  & $25\farcs8 \times 12\farcs5$ & $5\fdg5$ \\
2001-12-04  & $<420$ & $140$ & $32\farcs7 \times 11\farcs4$  & $-23\fdg9$ \\
2002-01-25  & $<225$ & $75$  & $0\farcs83 \times 0\farcs35$ & $-17\fdg6$ \\
2002-02-03  & $<240$ & $80$  & $0\farcs79 \times 0\farcs35$ & $-13\fdg1$ \\
2002-02-27  & $<249$ & $83$  & $1\farcs06 \times 0\farcs38$ & $-24\fdg8$ \\
2002-03-22  & $<237$ & $79$  & $0\farcs76 \times 0\farcs35$ & $-8\fdg9$ \\
2002-03-30  & $<273$ & $91$  & $0\farcs78 \times 0\farcs36$ & $-13\fdg9$ \\
2002-07-12  & $276$ & $68$  & $2\farcs44 \times 1\farcs13$ & $8\fdg5$ \\
2002-07-29  & $446$ & $53$  & $2\farcs71 \times 1\farcs12$ & $19\fdg3$ \\
2002-08-05  & $383$ & $60$  & $2\farcs35 \times 1\farcs11$ & $3\fdg9$ \\
2002-08-13  & $448$ & $57$  & $2\farcs35 \times 1\farcs11$ & $-3\fdg0$ \\
2002-09-18  & $236$ & $58$  & $8\farcs76 \times 1\farcs97$ & $50\fdg8$ \\
2002-12-07  & $425$ & $52$  & $12\farcs3 \times 3\farcs61$ & $33\fdg0$ \\
2003-01-18  & $462$ & $83$  & $36\farcs6 \times 5\farcs61$ & $49\fdg0$ \\
2003-02-06  & $370$ & $74$  & $27\farcs6 \times 11\farcs2$ & $-19\fdg3$ \\
2003-02-26  & $370$ & $74$  & $30\farcs5 \times 13\farcs1$ & $-22\fdg6$ \\
2003-07-31  & $426$ & $64$  & $1\farcs58 \times 0\farcs39$ & $41\fdg5$ \\
2003-10-27  & $431$ & $54$  & $3\farcs27 \times 1\farcs10$ & $28\fdg4$ \\
2003-11-21  & $402$ & $57$  & $4\farcs97 \times 1\farcs11$ & $39\fdg1$ \\
[5pt]
\hline 
\end{tabular} 
\end{center}
\caption{Log of the VLA radio core observations at 4.9 GHz, between 2001--2003.
The data for this table were obtained from images available in the NRAO VLA Archive Survey database.
Upper limits are $3\sigma$ levels.}
\label{tab1}
\end{table}

\begin{table}
\begin{center}
\begin{tabular}{lrrr}\hline
\hline
Date & Instrument & Mode & Exp time\\
\hline\\[-5pt]
2000-09-19 & pn & Small Window  & 8.97 ks\\
   & MOS1 & Timing & 9.87 ks\\
   & MOS2 & Partial RFS & 1.33 ks\\[5pt]
2001-03-22 & pn & Large Window  & 18.43 ks\\
   & MOS1 & Partial W2 & 20.95 ks\\
   & MOS2 & Partial W2 & 20.95 ks\\
   & RGS1+2 & Spectro+Q & 2 $\times$ 18.3 ks \\[5pt]
2002-09-28 & pn & Small Window  & 5.94 ks\\
   & MOS1 & Timing & 8.28 ks\\
   & MOS2 & Partial W2 & 8.36 ks\\
   & RGS1+2 & Spectro+Q & 2 $\times$ 8.53 ks \\
\hline
2008-04-01 & XRT & PC & 6.9 ks  \\
 &  & WT & 0.45 ks \\
2008-04-06 & XRT & PC & 3.3 ks\\
 &  & WT  & 0.38 ks\\
2008-04-07 & XRT & PC & 3.5 ks\\
 &  & WT  & 0.23 ks\\
2008-04-11 & XRT & PC & 7.4 ks\\
 &  & WT & 0.61 ks\\[5pt]
2009-03-12 & XRT & WT  & 2.08 ks\\
2009-03-13 & XRT & WT  & 5.04 ks\\
\hline 
\end{tabular} 
\end{center}
\caption{Log of the {\it XMM-Newton}/EPIC+RGS observations (top) and 
{\it Swift}/XRT observations (bottom) used in this paper.}
\label{tab2}
\end{table}



\section{Radio observations}

\subsection{VLA data from 2001--2003}

We searched the NRAO VLA Archive Survey database\footnote{http://www.aoc.nrao.edu/$\sim$vlbacald/~} 
for observations of GRS 1758$-$258 during the 2000--2003 sequence of state transitions. 
A total of 38 unpublished, processed radio maps at 4.86 GHz were found covering 
the period 2001 February 1 -- 2003 November 21 (Table 1).
The median root-mean-square (rms) noise level near GRS 1758$-$258 
is about 75 $\mu$Jy beam$^{-1}$. 
We usually found that the best way to determine the core flux density 
was either to fit a point-source elliptical Gaussian, or to use 
the flux density of the peak pixel in the map. However, for the maps 
from 2003 February 6 and 26, where it appears that the core is extended, 
we instead summed the pixels in a box or polygon surrounding the component.
The angular resolution was generally too fine to obtain lobe flux measurements 
(the extended emission was resolved out), and the few maps with lower angular resolution 
did not have enough sensitivity; therefore, we only used the VLA maps for studying 
the core emission.

\subsection{ATCA observations from 2008--2009}

We observed GRS 1758$-$258 at 6 cm with the ATCA in four separate sessions 
over the period 2008 December 29 -- 2009 January 02. The array configuration 
was 6C, with minimum and maximum baselines of 153 and 6000 m, respectively. 
We observed simultaneously in two bands centred at 4800 and 5423 MHz, 
each with an effective bandwidth of 104 MHz. The total integration time 
on-source was 35.03 h, spread approximately evenly across the four sessions. 
The primary calibrator was B1934$-$638, and B1752$-$225 the secondary calibrator.               

Data reduction and imaging was carried out with {\sc miriad} \citep[][]{sau95}. 
We imaged all of the data together using a robust weighting parameter 
\citep[][]{bri95} of 0.5; the effective frequency of the combined data 
is 5.09 GHz. In addition, the multi-frequency deconvolution task {\it mfclean} 
\citep[][]{sau94} was used because of the frequency gap between the two bands. 
The cleaned, primary-beam-corrected image (Figure 3) has an angular resolution 
of $5\farcs08 \times 1\farcs69$ (beam position angle $1\fdg9$). 
We also tapered the data to produce an image with the same resolution 
as the map made from the combined 1992 and 1997 VLA data 
(Fig. 3 in \citealt[][]{mar02}); the angular resolution in this case 
is $10\farcs1 \times 6\farcs0$ (beam position angle $9\fdg7$). 
A tapered map provides better sensitivity to the low-surface-brightness 
extended emission from the lobes, which is resolved out in our higher-resolution image. 
In the vicinity of GRS 1758$-$258, the rms noise levels in the two maps 
are 16 $\mu$Jy beam$^{-1}$ (higher resolution) and 26 $\mu$Jy beam$^{-1}$ (tapered). 
Moreover, we estimate that the calibration uncertainty relative 
to B1934$-$638 is about 5 per cent in both cases.


\begin{figure}\ \ 
\begin{center}
\psfig{figure=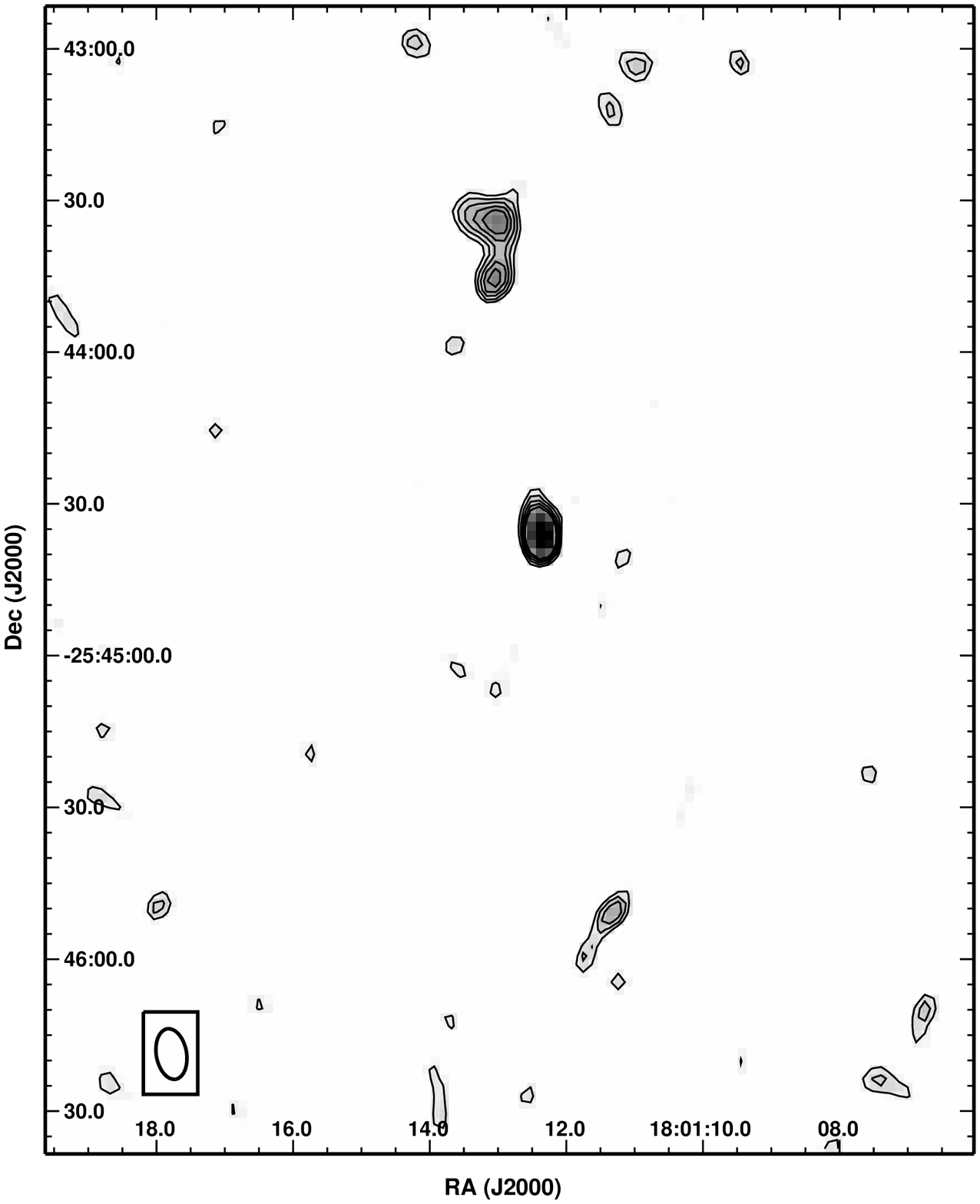,width=85mm,angle=0}\\
\psfig{figure=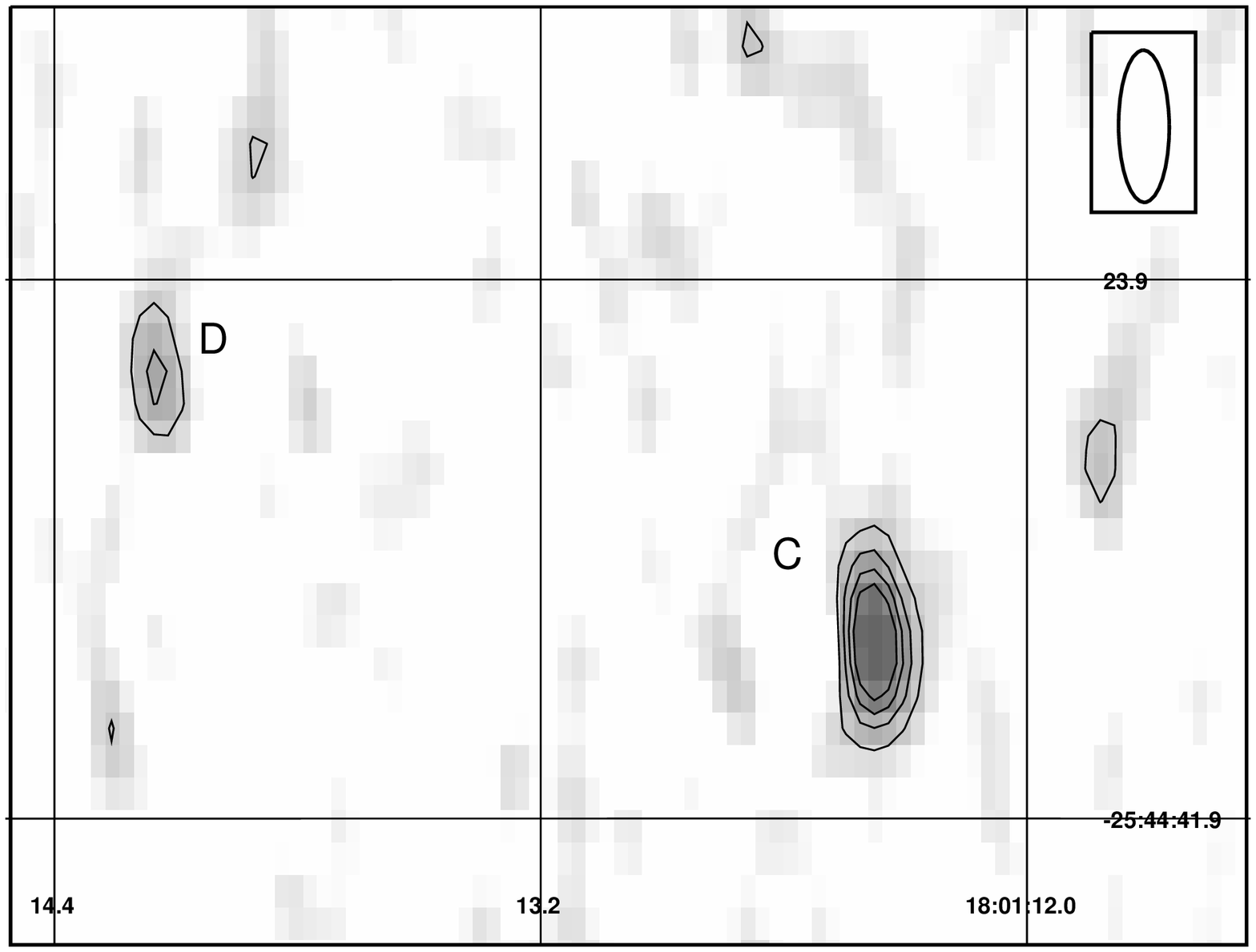,width=84mm,angle=0}
\end{center}
\caption{Top panel: tapered radio map at 6 cm (5.09 GHz), 
from our ATCA observations on 2008 December 29--2009 January 2.
Contour levels are 65, 78, 91, 104, 107 $\mu$Jy beam$^{-1}$, corresponding 
to 2.5, 3, 3.5, 4, 4.5 times the rms noise. 
The beam size is $10\farcs1 \times 6\farcs0$, and its position angle is $9\fdg7$.
Bottom panel: higher-resolution zoomed-in view of the core region 
(source C) from the same ATCA data. 
Contour levels are 48, 80, 112, 144 $\mu$Jy beam$^{-1}$, corresponding 
to 3, 5, 7, 9 times the rms noise. 
The beam size is $5\farcs09 \times 1\farcs69$, and its position angle is $1\fdg9$.
Source D is probably a background source, see \citet{mar02,har05}.}
\label{f3}
\end{figure}

\section{Main results of our radio study}

\subsection{Radio core behaviour during 2001--2003}

The behaviour of the radio core during the 2001 soft state transition 
mirrored that observed during the previous major soft transition 
in 1992 \citep{gil93,gre97,kec01}.
Between 1992 January and July, the core was undetected at 5 GHz, 
at a flux density $< 0.07$ mJy \citep{mir94}. It was detected again  
on 1992 September 10--26 at $0.53 \pm 0.04$ mJy \citep{mir94} 
as the X-ray source returned to a harder state. It remained visible 
and varied between $\approx 0.2$--$0.3$ mJy during 1993 \citep{mir94}, 
and between $\approx 0.1$--$0.2$ mJy in 1997 \citep{lin00,mar02,har05}, 
while the X-ray source persisted in a hard state. 

Systematic VLA monitoring started again in 2001 February (Table 1), 
when the core was not detected, down to a limit of $0.24$ mJy (3 times the rms noise). 
At that epoch, the X-ray source was undergoing a series of rapid spectral 
state changes \citep{smi01a}. The short transition timescale suggests 
that there was already a full disc, partly covered by a rapidly varying 
Comptonizing region (as discussed in Sections 6.1, 7.2, and in \citealt{smi01a}).
At the end of 2001 February, GRS 1758$-$258 settled into a steadier 
soft state \citep{smi01a}. The radio core continued to be undetected 
during that month, down to $\la 0.1$ mJy (Table 1). 
It was undetected throughout the rest of 2001, although the VLA observations 
were too short to place very stringent constraints. 
When the {\it RXTE}/PCA X-ray monitoring resumed on 2002 February 5, 
the $2$--$10$ keV count rate had already increased to a value close 
to its long-term average (Figure 1), consistent with a return to 
an intermediate or hard X-ray state. However, the radio core was still 
undetected to $\la 0.24$ mJy until at least 2002 March 30 (Table 1).
The radio jet finally appeared in the 2002 July 12 snapshot, at 
a 5-GHz flux density $\approx 0.27$ mJy. It was then detected in every 
subsequent observation from 2002 and 2003
(peak pixel $> 3.5\sigma$ in all cases), varying 
between $\approx 0.24$--$0.46$ mJy (Table 1).

Overall, the radio core behaviour is in agreement with the theoretical 
expectations and with the behaviour of other Galactic BH transients 
\citep{cor01,fen04,bel10}: 
the jet is suppressed when the X-ray source is in 
a disk-dominated thermal state. It appears that the jet 
had already vanished or at least strongly declined in the intermediate 
states immediately before and after the soft state (2001 February and 
2002 January--March).
However, the sparse sampling and shallow VLA detection limits 
do not allow us to constrain the transition more strongly.

\subsection{Radio core and lobes in 2008--2009}

We found core emission centred at 
RA(J2000) $ = 18^{\rm h}01^{\rm m}12\fs39$, 
Dec(J2000) $= -25^{\circ} 44\arcmin 36\farcs0$ (Figure 1), 
with an error of $0\farcs08$ in RA and $0\farcs2$ in Dec.
This is still consistent with the position determined 
from the highest-resolution VLA observations \citep[][and 
from our direct re-analysis of the VLA data]{mun10}.
In the higher-resolution image, the integrated core flux density, 
determined from an elliptical Gaussian fit with the {\sc miriad} task {\it imfit}, 
is $0.25 \pm 0.02$ mJy. Similarly, the integrated core flux density 
in the tapered map is $0.25 \pm 0.03$ mJy. Though there is a hint 
of source extension for the core component in the higher-resolution map 
(Figure 3, bottom panel), more sensitive data are required to confirm this. 
The location of the two extended lobes is also consistent 
with those mapped in the 1997 observations \citep{mar02,har05}.
The peak flux density in the northern lobe (also known 
as VLA source B) is $135 \mu$Jy beam$^{-1}$ at 
RA(J2000) $ = 18^{\rm h}01^{\rm m}13\fs04$, 
Dec(J2000) $= -25^{\circ} 43\arcmin 33\farcs9$. 
Its integrated flux density is $(0.4\pm 0.1)$ mJy, 
which we determined by summing the pixels in a box surrounding the lobe.
The peak flux density in the southern lobe (VLA source A) 
is $103 \mu$Jy beam$^{-1}$ at 
RA(J2000) $ = 18^{\rm h}01^{\rm m}11\fs37$, 
Dec(J2000) $= -25^{\circ} 45\arcmin 50\farcs8$; 
we cannot get a reliable integrated flux density 
because the signal-to-noise level is too low.
We do not have enough sensitivity in the tapered image  
to determine whether the lobes have changed in flux 
and morphology. 
The integrated flux density for the northern lobe 
was listed as $(0.52\pm 0.05)$ mJy in 1992 \citep{rod92} 
and $(0.65\pm 0.06)$ mJy in 1997 \citep{mar02}. 
We also note that the nearby VLA source D, located $\approx 25\arcsec$ 
north-east of GRS 1758$-$258, well visible 
in 1992 and 1997 \citep{rod92,mar02,har05} and interpreted 
as a background source, has faded somewhat; its peak flux density
in the higher-resolution image is $\approx 0.09$ mJy beam$^{-1}$. 


\begin{figure}
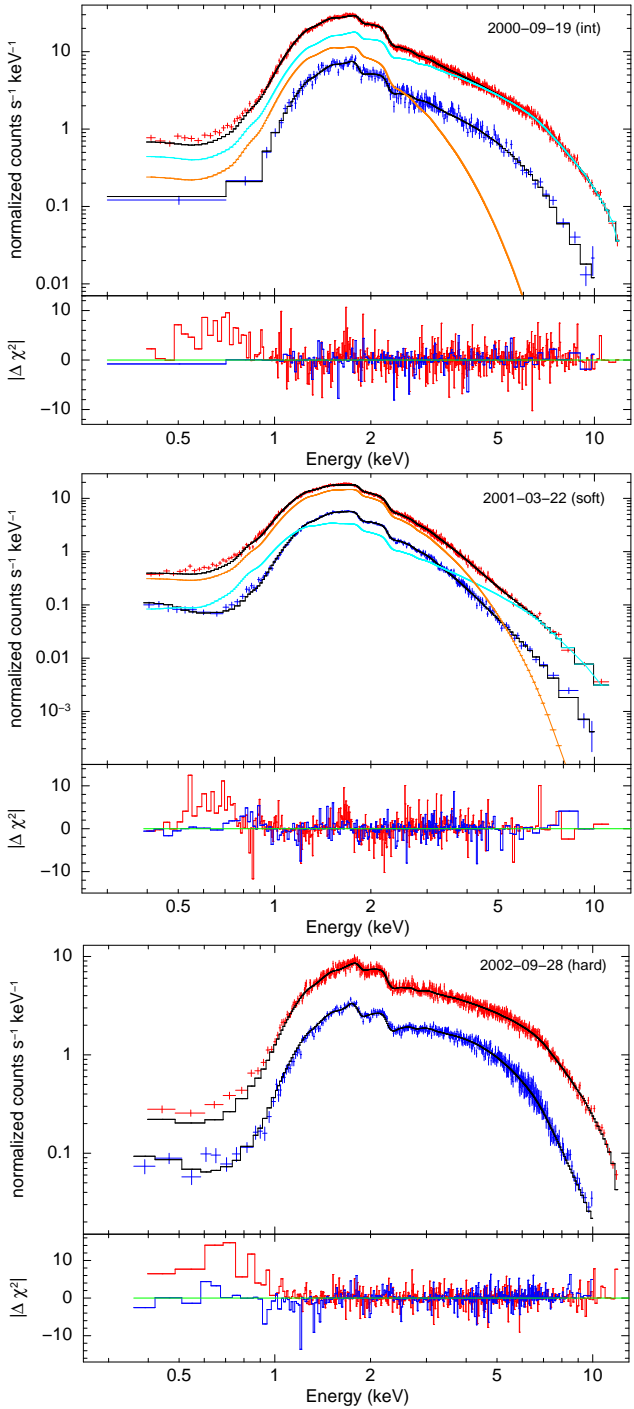

\begin{center}
\psfig{figure=2000_pnmos2_rev.ps,width=62mm,angle=270}\\
\psfig{figure=2001_pnmos_rev.ps,width=62mm,angle=270}\\
\psfig{figure=2002_pnmos2_rev.ps,width=62mm,angle=270}
\end{center}
\caption{Top panel: {\it XMM-Newton}/EPIC spectra from the 2000 observation, 
fitted with an absorbed disk-blackbody plus power-law model; 
in the online colour figures, pn data are in red, MOS2 in blue. 
The cyan and orange curves represent the power-law 
and disk-blackbody components in the best fitting pn model; 
see Tables 3,4 for the fit parameters. 
Middle panel: {\it XMM-Newton}/EPIC spectra from the 2001 observation, 
fitted with an absorbed disk-blackbody plus power-law model; 
pn data are in red, MOS2 in blue in the online version. 
The cyan and orange curves represent the power-law 
and disk-blackbody components in the best fitting pn model (Table 5). 
Bottom panel: {\it XMM-Newton}/EPIC spectra from the 2002 observation, 
fitted with an absorbed broken power-law model (Table 6); 
pn data are in red, MOS2 in blue in the online version.}
\label{f4}
\end{figure}

\begin{figure}
\begin{center}
\psfig{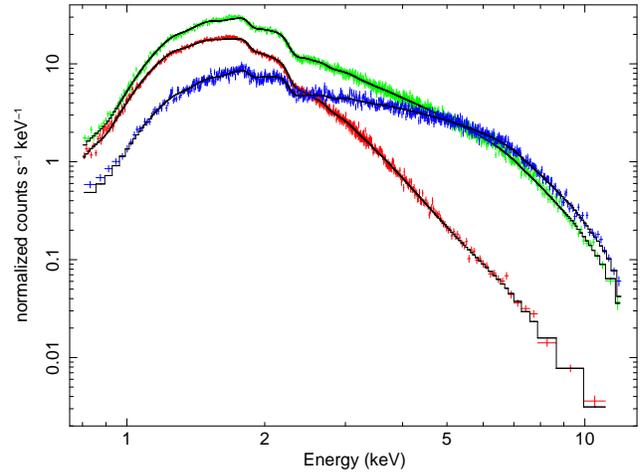}\\
\end{center}
\caption{
Comparison between {\it XMM-Newton}/EPIC-pn spectra 
in 2000 (green in the online version; intermediate state), 2001 (red; soft state) 
and 2002 (blue; hard state).}
\label{f5}
\end{figure}

\begin{figure}
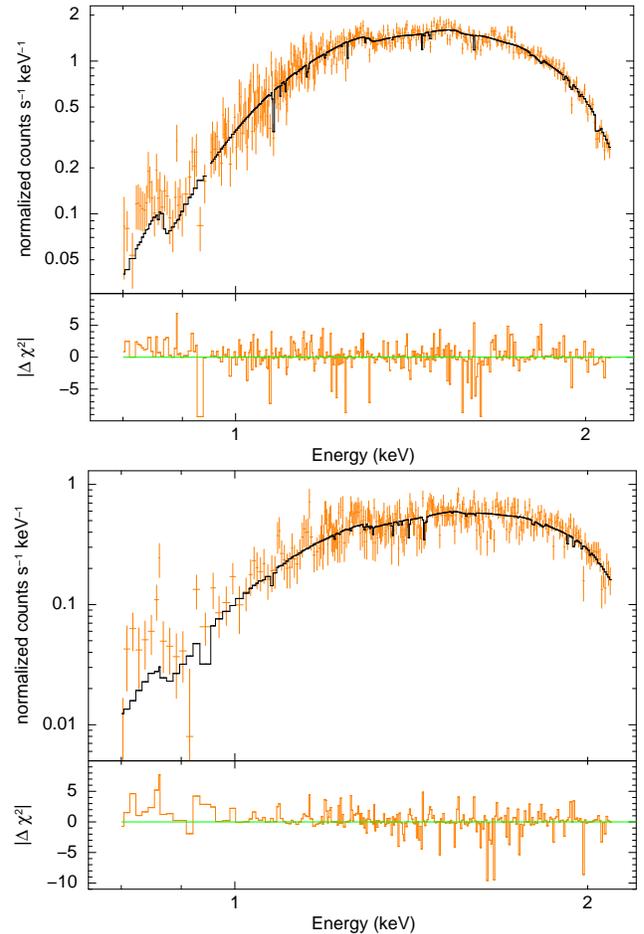

\begin{center}
\psfig{figure=2001_rgs_diskbbpo.ps,width=62mm,angle=270}\\
\psfig{figure=2002_rgs_po.ps,width=62mm,angle=270}
\end{center}
\caption{Top panel: {\it XMM-Newton}/RGS spectrum from 2001, 
fitted with an absorbed disk-blackbody plus power-law model. 
Bottom panel: {\it XMM-Newton}/RGS spectrum from 2002, 
fitted with an absorbed power-law model.
See Table 7 for the best-fitting parameters.}
\label{f6}
\end{figure}

\section{X-ray observations}

\subsection{{\it XMM-Newton} observations from 2000--2002}

GRS 1758$-$258 was observed with {\it XMM-Newton} on 
2000 September 19, 2001 March 22 and 2002 September 28 (Table 2): 
that is, immediately before, during and immediately after 
the 2001 soft state.
An analysis of the 2001 Reflection Grating Spectrograph (RGS) data was presented 
by \citet{mil02}. Here we present the results 
of the European Photon Imaging Camera (EPIC) observations in the three epochs, 
and of the 2002 RGS observation; we also re-examine the RGS spectrum from 2001 
in view of the new EPIC results.

We downloaded the Observation Data Files from the public archive 
and used the Science Analysis System ({\footnotesize SAS}) version 10.0.0 (xmmsas\_20100423)
to process and filter the event files and extract spectra and lightcurves. 
We selected single and double events (pattern $\le 4$ for the pn and pattern 
$\le 12$ for the MOS), with the standard flagging criteria \#XMMEA\_EP and 
\#XMMEA\_EM for the pn and MOS respectively. 
After building response and ancillary response files 
with the {\footnotesize SAS} tasks {\it rmfgen} and {\it arfgen}, 
we used {\footnotesize XSPEC} \citep{arn96} version 12 for spectral fitting,
and standard {\footnotesize FTOOLS} tasks for timing analysis.

The observing modes for the three EPIC instruments reflect attempts 
to reduce a potential pile up problem.
In 2000, the pn camera was in small window mode, MOS1 in timing mode, 
and MOS2 in the non-standard prime partial refresh frame store (RFS) mode. In 2001, 
the pn was in large window mode, and both MOSs in partial window mode.
In 2002, the pn was in small window mode, MOS1 in timing mode, 
and MOS2 in partial window mode.
The EPIC data are clearly affected by pile up: typical pn count rates 
in a $45\arcsec$ circle were $\approx 100$ ct s$^{-1}$ and $\approx 40$ ct s$^{-1}$ 
for the MOS in the first two observations, and about half that value 
in the third observations. We expect a deterioration in the response 
for count rates $\ga 50$ ct s$^{-1}$ for the pn in small window mode, 
and $\ga 6$ ct s$^{-1}$ in large window mode; pile up affects the MOS 
for rates $\ga 5$ ct s$^{-1}$ in small window mode. 
The MOS1 observations in timing mode from 2000 and 2002 
do not suffer from pile-up. However, using timing data for spectral 
analysis introduces its own set of calibration uncertainties, 
that are probably as significant as the pile-up error.

We used the {\footnotesize SAS} task {\it epatplot} to estimate 
the level of pile up for different source extraction regions, by comparing 
the observed and expected distribution of single and multiple events.
After several tests, we chose to extract the source spectra from annuli 
between an inner radius of $10\arcsec$ and an outer radius of $45\arcsec$.
We repeated the extraction and spectral analysis 
for source annuli with inner radii of $5\arcsec$, $15\arcsec$ 
and $20\arcsec$ and found that the difference in the fit parameters 
were within the error bars of the values for a $10\arcsec$ inner radius.
Thus, we cannot completely eliminate the errors due 
to pile-up, but we are confident that they are a small effect 
compared with the spectral state transitions discussed in this paper. 

Another calibration issue we had to deal with 
is the use of the RFS read-out mode for the 2000 MOS2 
observation. This mode has been used only for few observations 
with high count rates ({\it e.g.}, the Crab Nebula and 
the N132D supernova remnant) at very early stages 
of the {\it XMM-Newton} mission. The flux in this observing mode 
has never been properly calibrated. After consultations with 
the calibration scientists, and from a comparison with those early 
observations, we concluded that the RFS mode underestimates 
the true flux by about 30\%, but the spectral shape 
is mostly reliable, consistent with spectra obtained in other modes.
To double-check that, we fitted the 2000 MOS2 spectra twice: 
once independently, and once with fit parameters fixed equal to the pn 
parameters but with a free normalization constant (which turned out 
to be $\approx 0.7$, as expected).

For the pn observation in large window mode, we extracted the background 
from three $45\arcsec$ circles at sufficiently large distances from the source. 
For the 2000 and 2002 pn observations, in small window mode, 
we extracted the background from 
two circular regions of $45\arcsec$ radius, located at the bottom 
of the observing window, $\approx 3\arcmin$ from the source position.
However, we checked that the background count rate is only 
modestly significant for the 2000 pn observation, when 
it is $\sim 2\%$ of the source count rate, and slightly affects 
the spectral fit only at energies $\ga 10$ keV. In 2001 and 2002, 
the background count rate is $< 1\%$ of the source rate, and 
its effect on the spectral shape is much less than 
the errors due to calibration uncertainties and pile-up 
corrections. For the MOS observations, we could not extract 
a background from the central chip because the whole observing 
window is affected by the source emission. From the count rate 
in the outer chips, we estimated that the MOS background rate 
in our fitting band is $< 1\%$ in all three epochs, 
and can be safely ignored.

For all spectral fits, we rebinned the spectra to a minimum 
of 30 counts per bin, so that we could use Gaussian statistics.
For plotting purposes only, we rebinned the data to a minimum 
signal-to-noise ratio of 12 for the pn and 6 for the MOS spectra.  
We fitted the pn spectra in the $0.8$--$12$ keV range, and 
the MOS spectra in the $0.8$--$10$ keV range. We plotted 
the spectra down to $0.4$ keV in Figure 4 for illustration 
purposes only. The small discrepancy between pn and MOS at 
$0.4$--$0.8$ keV may be due to a number of reasons: a different 
level of pile-up, a higher contamination from extended thermal 
emission or nearby soft sources in the pn (which has a larger 
point spread function), a more complex structure of the absorbing 
medium, or cross-calibration uncertainties. 
A soft excess below $0.9$ keV is also clearly visible 
in the RGS spectra from 2001 \citep{mil02}.
In any case, these are minor uncertainties, irrelevant 
to the main objectives of our spectral study.

The {\it XMM-Newton}/RGS data were also processed and analysed from 
the Observation Data Files with the {\sc sas} 10.0 software. 
Both RGS detectors were operated in the standard Spectro+Q mode 
for the 2001 and 2002 observations. (Unfortunately, during 
the 2000 observation both RGS1 and RGS2 were inactive 
and thus no RGS data were obtained.) 
We filtered out time intervals with count rates $>0.3$ ct s$^{-1}$ 
in CCD number 9, which is the most affected by background flares. 
For the 2001 observation, we removed $\approx 3$ ks, leaving a net exposure time 
of $\approx 18.3$ ks for each RGS detector. For the 2002 observation, 
the background level was lower and no time filtering was required; 
the net exposure time is $\approx 8.5$ ks. 
After extracting the source and background spectra, 
we created response matrices with {\it rgsproc}. 
For the purpose of spectral fitting, we combined the first-order spectra 
and response matrices of RGS1 and RGS2 for each observation, using the SAS task 
{\it rgscombine}. The channels in the RGS spectra were then rebinned 
by a factor of 2 to increase the number of counts per bin. 
We used the $0.8$--$2.1$ keV energy range ($10$--$26$ \AA) for spectral 
fitting in {\sc xspec}.

\subsection{{\it Swift}/XRT observations in 2008--2009}

We want to relate our ATCA radio observations of 2008--2009 
to the X-ray spectral state at the time. To do that, 
we looked for archival X-ray observations as close as possible 
to that epoch. {\it Swift}/XRT observed the source 
four times between 2008 April 01--11, and twice on 
2009 March 12--13 (Table 2; 
see NASA's Heasarc data archive for a more detailed logbook). 
We used the on-line XRT data product generator
\citep{eva09} to extract lightcurves and spectra, including 
background and ancillary response files; we selected grade 0--12 events. 
We downloaded the suitable spectral response files for single and double events 
in photon-counting (PC) mode and window-timing (WT) mode from the latest Calibration Database 
(2009 December 1).

In both the 2008 and 2009 observations, the XRT count rate is high 
enough (a few counts s$^{-1}$) to create serious pile-up problems 
in PC mode, but not in WT mode. To attenuate the pile-up problem, 
the source extraction procedure \citep{eva09} first determines whether the count rate 
within a 30 pixel ($70\farcs8$) radius is $> 0.6$ counts s$^{-1}$: that is the level 
at which the observed single-pixel event rate is $\approx 90\%$ 
of the input single-pixel event rate \citep{vau06}. Then, 
a point-spread-function profile is obtained and compared 
with a default non-piled-up profile. From this, the extraction 
procedure determines the radius within which the two profiles 
differ significantly. Finally, the source extraction region is 
defined as the annulus between that radius and $70\farcs8$.

The WT-mode spectra do not suffer from pile up (it would only 
be significant at $\ga 100$ counts s$^{-1}$). However, 
the lack of spatial information does not allow background 
subtraction. That appears to be a problem at low energies 
($< 1$ keV), for the particle induced background.
In fact, at every epoch when both PC and WT spectra are available, 
the latter show a low-energy excess with respect to the PC, 
which is most likely due to soft proton contamination rather 
than true emission from the source. At energies $\ga 1$ keV, 
PC and WT spectra agree within the observational error.

We examined the four individual spectra from the 2008 and 
the two spectra from 2009, and found that the multiple spectra
from each epoch were consistent with each other. The {\it Swift}/BAT 
lightcurve (Figure 2) confirms that there were no unsual flares 
or state transitions during the two epochs. We coadded the spectra 
from each of the two epochs to increase the signal to noise. 
We rebinned and fitted the coadded spectra with {\footnotesize{XSPEC}} Version 12.

\begin{figure}[!]
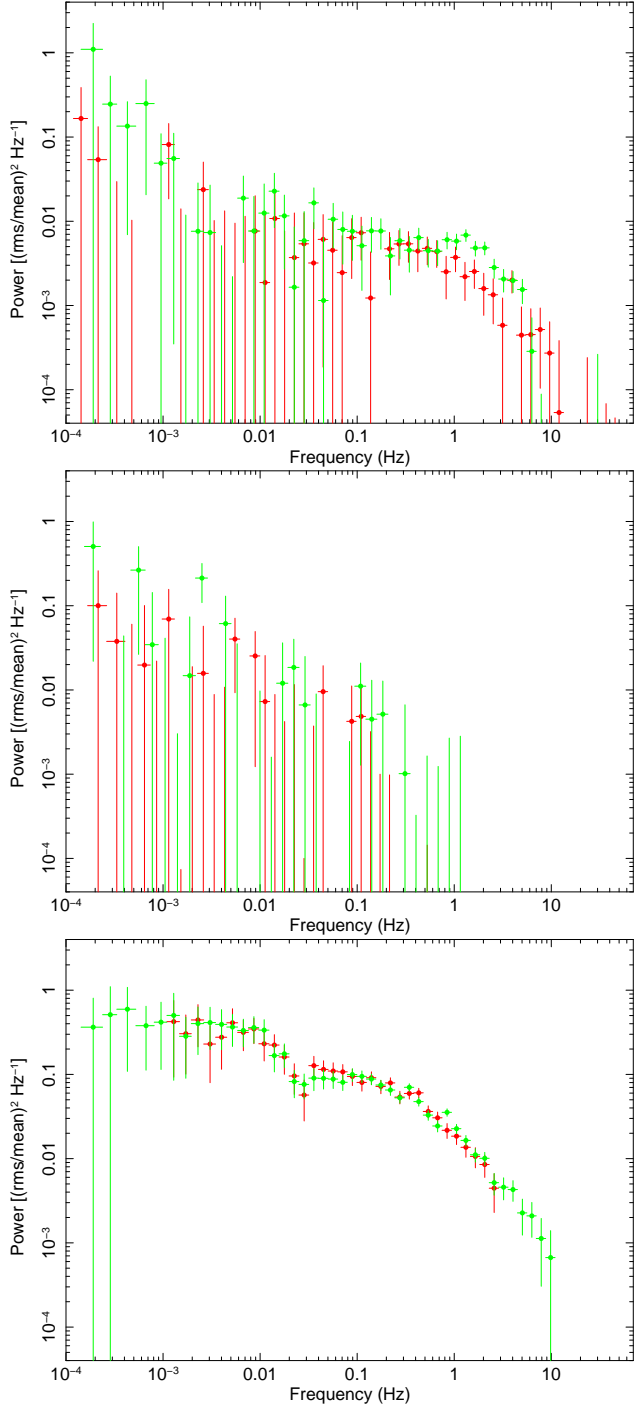

\begin{center}
\psfig{figure=2000_pds.ps,width=62mm,angle=270}\\
\psfig{figure=2001_pds.ps,width=62mm,angle=270}\\
\psfig{figure=2002_pds.ps,width=62mm,angle=270}
\end{center}
\caption{Power density spectra in 2000, 2001 and 2002. In each epoch, 
the red datapoints (shown in colour in the online version) 
are the power density spectra extracted 
from the EPIC pn lightcurves; the green datapoints are extracted from the MOS1 data.
The three spectra are typical of the intermediate state, high/soft state, 
and low/hard state respectively.}
\label{f7}
\end{figure}

\begin{table}
\begin{center}
\begin{tabular}{lrr}
\hline
Parameter & {pn Value} & {MOS2 Value} \\
\hline
\multicolumn{3}{c}{Model: {\tt phabs*phabs*(diskbb+pow)}} \\
\hline\\[-5pt]
$N_{H,\rm Gal}$ & $0.75$ & $0.75$ \\[5pt]
$N_{H,\rm int}$ & $0.79^{+0.02}_{-0.02}$ 
         & $0.91^{+0.12}_{-0.10}$\\[5pt]
$kT_{\rm dbb}$ 
        & $0.48^{+0.01}_{-0.02}$ & $0.48^{+0.06}_{-0.06}$\\[5pt]
$N_{\rm dbb}$ 
        & $693^{+93}_{-79}$ & $683^{+577}_{-265}$\\[5pt]
$\Gamma$ & $2.24^{+0.03}_{-0.03}$
         & $2.04^{+0.24}_{-0.27}$\\[5pt]
$N_{\rm po}$ & $3.3^{+0.3}_{-0.1}$
         & $1.7^{+0.9}_{-0.6}$ \\[5pt]
\hline\\[-5pt]
$f_{0.3-12}$ & $7.6^{+0.1}_{-0.1}$ 
        & ($5.5^{+0.1}_{-0.4}$)\\[5pt]
$L_{0.3-12}$ & $1.9^{+0.1}_{-0.1}$ 
        & ($1.3^{+0.3}_{-0.2}$)\\[5pt]
$L_{0.3-100}$ & $2.3^{+0.1}_{-0.1}$ 
        & ($1.7^{+0.3}_{-0.2}$)\\[5pt]
$L^{\rm dbb}_{\rm bol}$ & $0.6^{+0.1}_{-0.1}$ 
        & $0.6^{+0.1}_{-0.1}$\\[5pt]
$F^{\rm dbb}_{0.3-12}$ & $27\%$ 
        & $40\%$\\[5pt]
\hline\\[-5pt]
$\chi^2_{\nu}$ &  $1.08~(1781.8/1658)$ & $1.03~(252.4/246)$ \\[5pt]
\hline 
\end{tabular} 
\end{center}
\caption{Best-fitting spectral parameters for the 2000 September 19 
{\it XMM-Newton}/EPIC observation.
Units: $N_{H,\rm Gal}$ (line-of-sight Galactic extinction, from \citealt{kal05})
and $N_{H,\rm int}$ (intrinsic extinction) are in units of $10^{22}$ cm$^{-2}$; 
$kT_{\rm dbb}$ in keV; $N_{\rm po}$ in $10^{-1}$ photons keV$^{-1}$ cm$^{-2}$ 
s$^{-1}$ at 1 keV; the observed flux is in $10^{-10}$ erg cm$^{-2}$ s$^{-1}$; 
the emitted luminosities in $10^{37}$ erg s$^{-1}$. $F^{\rm dbb}_{0.3-12}$ 
is the disk-blackbody fraction of emitted luminosity in the $0.3$--$12$ keV band. 
Errors indicate the 90\% confidence interval for each parameter of interest. 
As explained in the text (Section 5.1), the MOS2 flux values are $\approx 30\%$
lower than the more reliable pn values.}
\label{tab3}
\end{table}

\begin{table}
\begin{center}
\begin{tabular}{lrr}
\hline
Parameter & {pn Value} & {MOS2 Value} \\
\hline
\multicolumn{3}{c}{Model: {\tt const*phabs*phabs*(diskbb+pow)}} \\
\hline\\[-5pt]
Const & $1.00$ & $0.69^{+0.01}_{-0.01}$\\[5pt]
$N_{H,\rm Gal}$ & \multicolumn{2}{c}{$0.75$}  \\[5pt]
$N_{H,\rm int}$ & \multicolumn{2}{c}{$0.80^{+0.02}_{-0.02}$}\\[5pt]
$kT_{\rm dbb}$ 
        & \multicolumn{2}{c}{$0.48^{+0.01}_{-0.02}$} \\[5pt]
$N_{\rm dbb}$ 
        & \multicolumn{2}{c}{$705^{+93}_{-79}$} \\[5pt]
$\Gamma$ & \multicolumn{2}{c}{$2.24^{+0.03}_{-0.03}$}\\[5pt]
$N_{\rm po}$ & \multicolumn{2}{c}{$3.4^{+0.02}_{-0.02}$}\\[5pt]
\hline\\[-5pt]
$f_{0.3-12}$ & \multicolumn{2}{c}{$7.6^{+0.1}_{-0.1}$}\\[5pt]
$L_{0.3-12}$ & \multicolumn{2}{c}{$1.9^{+0.1}_{-0.1}$} \\[5pt]
$L_{0.3-100}$ & \multicolumn{2}{c}{$2.3^{+0.1}_{-0.1}$} \\[5pt]
$L^{\rm dbb}_{\rm bol}$ & \multicolumn{2}{c}{$0.6^{+0.1}_{-0.1}$}\\[5pt]
$F^{\rm dbb}_{0.3-12}$ & $27\%$ 
        & $27\%$\\[5pt]
\hline\\[-5pt]
$\chi^2_{\nu}$ &  \multicolumn{2}{c}{$1.06~(2091.7/1980)$} \\[5pt]
\hline 
\end{tabular} 
\end{center}
\caption{Best-fitting spectral parameters for the 2000 September 19 
{\it XMM-Newton}/EPIC observation, using a simple scaling constant 
between pn and MOS2 to account for the known calibration discrepancy (Section 5.1). 
Units are as in Table 3.
Errors indicate the 90\% confidence interval for each parameter of interest.}
\label{tab4}
\end{table}

\begin{table}
\begin{center}
\begin{tabular}{lrrr}
\hline
Parameter & {pn Value} & {MOS1 Value} & {MOS2 Value} \\
\hline
\multicolumn{4}{c}{Model: {\tt phabs*phabs*(diskbb+pow)}} \\
\hline\\[-5pt]
$N_{H,\rm Gal}$ & $0.75$ & $0.75$ & $0.75$\\[5pt]
$N_{H,\rm int}$ & $0.72^{+0.03}_{-0.03}$ 
         & $0.99^{+0.04}_{-0.02}$
         & $1.02^{+0.07}_{-0.04}$\\[5pt]
$kT_{\rm dbb}$ 
        & $0.49^{+0.01}_{-0.01}$ & $0.45^{+0.01}_{-0.01}$
        & $0.45^{+0.01}_{-0.01}$\\[5pt]
$N_{\rm dbb}$ 
        & $707^{+26}_{-23}$ & $1668^{+112}_{-105}$
        & $1379^{+109}_{-98}$ \\[5pt]
$\Gamma$ & $3.31^{+0.15}_{-0.16}$
         & $2.85^{+0.33}_{-0.32}$
         & $3.28^{+0.39}_{-0.38}$\\[5pt]
$N_{\rm po}$ & $1.06^{+0.35}_{-0.26}$
         & $0.54^{+0.43}_{-0.24}$
         & $0.89^{+0.85}_{-0.44}$ \\[5pt]
\hline\\[-5pt]
$f_{0.3-12}$ & $1.9^{+0.1}_{-0.1}$ 
        & $2.2^{+0.1}_{-0.1}$
        & $1.7^{+0.1}_{-0.1}$\\[5pt]
$L_{0.3-12}$ & $1.1^{+0.1}_{-0.1}$ 
        & $1.2^{+0.1}_{-0.1}$
        & $1.2^{+0.1}_{-0.1}$\\[5pt]
$L_{0.3-100}$ & $1.1^{+0.1}_{-0.1}$ 
        & $1.2^{+0.1}_{-0.1}$
        & $1.2^{+0.1}_{-0.1}$\\[5pt]
$L^{\rm dbb}_{\rm bol}$ & $0.7^{+0.1}_{-0.1}$ 
        & $1.2^{+0.1}_{-0.1}$
        & $0.9^{+0.1}_{-0.1}$\\[5pt]
$F^{\rm dbb}_{0.3-12}$ & $55\%$ 
        & $83\%$
        & $65\%$\\[5pt]
\hline\\[-5pt]
$\chi^2_{\nu}$ &  $1.11$ & $1.31$ & $1.19$ \\
               & $(1047.5/941)$ & $(402.4/306)$ & $(343.0/289)$\\[5pt]
\hline 
\end{tabular} 
\end{center}
\caption{Best-fitting spectral parameters for the 2001 March 22 
{\it XMM-Newton}/EPIC observation. Units are as in Table 3.
Errors indicate the 90\% confidence interval for each parameter of interest.}
\label{tab5}
\end{table}

\begin{table}
\begin{center}
\begin{tabular}{lrr}
\hline
Parameter & {pn Value} & {MOS2 Value} \\
\hline
\multicolumn{3}{c}{Model: {\tt phabs*phabs*pow}} \\
\hline\\[-5pt]
Const & $1.00$ & $1.025^{+0.008}_{-0.008}$\\[5pt]
$N_{H,\rm Gal}$ & \multicolumn{2}{c}{$0.75$}  \\[5pt]
$N_{H,\rm int}$ & \multicolumn{2}{c}{$0.71^{+0.02}_{-0.02}$}\\[5pt]
$\Gamma$ & \multicolumn{2}{c}{$1.52^{+0.01}_{-0.02}$}\\[5pt]
$N_{\rm po}$ & \multicolumn{2}{c}{$1.06^{+0.02}_{-0.02}$}\\[5pt]
\hline\\[-5pt]
$f_{0.3-12}$ & $6.5^{+0.1}_{-0.1}$ 
        & $6.7^{+0.1}_{-0.1}$\\[5pt]
$L_{0.3-12}$ & $0.75^{+0.02}_{-0.02}$ 
        & $0.77^{+0.03}_{-0.03}$\\[5pt]
\hline\\[-5pt]
$\chi^2_{\nu}$ &  \multicolumn{2}{c}{$1.10~(2227.3/2020)$}\\[5pt]
\hline 
\hline
\multicolumn{3}{c}{Model: {\tt const*phabs*phabs*(diskbb+pow)}} \\
\hline\\[-5pt]
Const & $1.00$ & $1.025^{+0.008}_{-0.008}$\\[5pt]
$N_{H,\rm Gal}$ & \multicolumn{2}{c}{$0.75$}  \\[5pt]
$N_{H,\rm int}$ & \multicolumn{2}{c}{$0.75^{+0.03}_{-0.03}$}\\[5pt]
$kT_{\rm dbb}$ 
        & \multicolumn{2}{c}{$0.081^{+0.033}_{-0.029}$} \\[5pt]
$N_{\rm dbb}$ 
        & \multicolumn{2}{c}{$(125\times 10^5)^{+6.3\times 10^9}_{-121\times10^5}$} \\[5pt]
$\Gamma$ & \multicolumn{2}{c}{$1.53^{+0.02}_{-0.01}$}\\[5pt]
$N_{\rm po}$ & \multicolumn{2}{c}{$1.09^{+0.03}_{-0.03}$}\\[5pt]
\hline\\[-5pt]
$f_{0.3-12}$ & $6.5^{+0.1}_{-0.1}$ 
        & $6.7^{+0.1}_{-0.1}$\\[5pt]
$L_{0.3-12}$ & $2.3^{+0.1}_{-0.1}$ 
        & $2.4^{+0.3}_{-0.2}$\\[5pt]
$L^{\rm po}_{0.3-12}$ & $0.75^{+0.02}_{-0.02}$ 
        & $0.77^{+0.03}_{-0.03}$\\[5pt]
\hline\\[-5pt]
$\chi^2_{\nu}$ &  \multicolumn{2}{c}{$1.09~(2205.4/2018)$}\\[5pt]
\hline 
\hline
\multicolumn{3}{c}{Model: {\tt const*phabs*phabs*bknpow}} \\
\hline\\[-5pt]
Const & $1.00$ & $1.023^{+0.008}_{-0.008}$\\[5pt]
$N_{H,\rm Gal}$ & \multicolumn{2}{c}{$0.75$}  \\[5pt]
$N_{H,\rm int}$ & \multicolumn{2}{c}{$0.68^{+0.02}_{-0.02}$}\\[5pt]
$\Gamma_1$
        & \multicolumn{2}{c}{$1.48^{+0.01}_{-0.01}$} \\[5pt]
$E_{\rm b}$ 
        & \multicolumn{2}{c}{$7.1^{+0.5}_{-0.4}$}\\[5pt]
$\Gamma_2$ & \multicolumn{2}{c}{$1.87^{+0.13}_{-0.10}$}\\[5pt]
$N_{\rm po}$ & \multicolumn{2}{c}{$1.02^{+0.02}_{-0.02}$}\\[5pt]
\hline\\[-5pt]
$f_{0.3-12}$ & $6.4^{+0.1}_{-0.1}$ 
        & $6.5^{+0.1}_{-0.1}$\\[5pt]
$L_{0.3-12}$ & $0.73^{+0.03}_{-0.03}$ 
        & $0.74^{+0.04}_{-0.04}$\\[5pt]
$L_{0.3-100}$ & $1.6^{+0.1}_{-0.1}$ 
        & $1.7^{+0.1}_{-0.1}$\\[5pt]
\hline\\[-5pt]
$\chi^2_{\nu}$ &  \multicolumn{2}{c}{$1.07~(2164.2/2018)$}\\[5pt]
\hline 
\end{tabular} 
\end{center}
\caption{Best-fitting spectral parameters for the 2002 September 28 
{\it XMM-Newton}/EPIC observation. Units are as in Table 3.
Errors indicate the 90\% confidence interval for each parameter of interest.}
\label{tab6}
\end{table}

\section{Main results of our X-ray study}

\subsection{X-ray spectral state in 2000--2002}

The main finding of our spectral analysis is that 
the source was in three different states during the three 
epochs: intermediate state (similar contributions from 
disk and power-law components) in 2000; soft state (disk dominated) 
in 2001; and hard state (power-law dominated) in 2002.
Here we will give a more detailed description of the spectral 
properties in the three states.
 
For the 2000 dataset, we used EPIC pn and MOS2 for the spectral 
analysis; we used MOS1 for the timing analysis (Section 6.2). 
We knew that the EPIC MOS2 calibration 
would underestimate the flux by $\approx 30\%$ (Section 5.1). 
We fitted the pn and MOS2 data in two different ways: first, with 
fully independent parameters, and then by locking the fit parameters 
together but allowing for a free normalization factor in the MOS2.
We find that single component models (either power-law 
or disk-blackbody) are ruled out, but an absorbed disk-blackbody 
plus power law model provides an acceptable fit. 
Overall, the pn and MOS parameters are consistent with each other 
(Tables 3,4), if we consider all the uncertainties outlined in Section 5.1.
The power-law photon index $\Gamma \approx 2.2$; the disk colour 
temperature $kT_{\rm in} \approx 0.48$ keV; the disk 
normalization $N_{\rm dbb} \equiv (r_{\rm in}/d_{10})^2 \cos \theta 
\approx 700$. Following a conventional procedure, for fitting 
purposes, we split the total column density into 
a fixed Galactic ``line-of-sight'' component \citep{kal05}\footnote{Although 
in this case we are only looking half way through the Galaxy, so the true 
line-of-sight extinction may be about half of that value.} and a free additional 
intrinsic component. The total column density $\approx 1.6 \times 10^{22}$ 
cm$^{-2}$ is similar to the values reported in previous X-ray 
studies. The unabsorbed luminosity in the $0.3$--$12$ keV band 
is $\approx 1.9 \times 10^{37}$ erg s$^{-1}$, and the direct disk 
emission contributes for $\approx 30\%$ in this band.

In 2001, all three EPIC instruments were set in spectral modes.
We fitted them independently and found that 
they give a consistent result (Table 5). The X-ray spectrum 
is again well fitted with an absorbed disk-blackbody plus 
power-law model. This time, the disk-blackbody component dominates, 
and the spectrum is consistent with a canonical high/soft state.
We are unable to remove the discrepancies between the three instruments 
in some of the fitting parameters. However, we need to take into 
account that the pn (large window mode) was more heavily piled up 
than the MOS---hence, a larger fraction of its soft flux is 
incorrectly shifted to higher energies (cf. $F^{\rm dbb}_{0.3-12}$ 
in Table 4). Furthermore, MOS2 had several columns, straddling 
the source extraction region, that were nominally unflagged but 
returned an anomalously low count rate compared with other regions 
symmetrically placed with respect to the source position.
Based on these considerations, we believe that MOS1 provides 
the most accurate spectral fit parameters in 2001. 
In summary, the power-law index $\Gamma \approx 2.9$, 
the disk colour temperature $kT_{\rm in} \approx 0.45$ keV, 
and the disk normalization $\approx 1700$.
The direct thermal emission from the disk 
contributes $\approx 80\%$ of the $0.3$--$12$ keV luminosity.
Luckily, there is a VLA observation taken almost simultaneously 
with the {\it XMM-Newton} observation, on 2001 March 22: 
the radio core was undetected down to $\approx 0.1$ mJy (Table 1), 
which confirms that the source was in the thermal-dominant state.

In 2002, we used EPIC pn and MOS2 for our spectral analysis.
The pile-up problem was somewhat less severe, and we found no other
instrumental problems. We fitted the data simultaneously, keeping 
the fit parameters constant apart from a free normalization factor 
(Table 6).
We found that a simple power-law with $\Gamma \approx 1.5$ is 
an acceptable fit. Adding a disk-blackbody component does not provide 
a significant improvement to the fit, and the disk temperature remains 
very low and essentially unconstrained (also because 
of the high column density). 
We obtain a moderate, statistically significant improvement 
($F$ statistic value $= 29.4$) by replacing the single power law 
with a broken power law (Table 6), with a slope $\Gamma \approx 1.5$ 
at $E \la 7$ keV, and $\Gamma \approx 1.9$ at higher energies.
Equally good fits are obtained using Comptonization models 
that account for the small degree of curvature in the spectrum.
The closest VLA observation to this epoch was taken on 2002 September 18, 
and shows radio flux from the core, at $\approx 0.24$ mJy (Table 1); this 
confirms that the jet had started again.

We tried to estimate a ``total'' X-ray luminosity by extrapolating  
the EPIC spectra to $0.3$--$100$ keV in the three epochs. We have strong 
evidence from other satellite observations 
\citep{gil93,kuz99,lin00,kec01,pot06} that the power-law spectrum 
extends to $\ga 100$ keV, at least in the hard state; on one occasion, 
it extended to $\approx 500$ keV \citep{loh11}.
The $0.3$--$100$ keV luminosity was 
$\approx 2.3 \times 10^{37}$ erg s$^{-1}$ in the intermediate state, 
$\approx 1.2 \times 10^{37}$ erg s$^{-1}$ in the soft state, and 
$\approx 1.6 \times 10^{37}$ erg s$^{-1}$ in the hard state.
In the hard state, part of the accretion power is transferred 
as kinetic energy to the electrons in the corona and/or jet. 
So, hard states should be less radiatively efficient than 
the thermal-dominant state. Given the inferred luminosities, 
it is very likely that the mass accretion rate $\dot{M}$ was also 
slightly higher in the hard state than in the soft state  
(unless the former was moderately beamed, see Section 7.2).
This is unusual for Galactic BHs, as previously noted \citep{smi01a,pot06}. 
Also unusual is the fact that over 20 years, the total X-ray luminosity 
has varied only by a factor of 3.

In all three epochs, the pn and MOS1 timing-mode spectra have 
a small excess at energies $\la 0.8$ keV, but that is not seen 
in the other MOS spectra, which have a higher spatial resolution 
and lower background contamination.
This residual emission is inconsistent with the highly absorbed 
blackbody plus power-law components used to model the spectrum 
at energies $> 0.8$ keV.
It can be formally modelled with an additional optically-thin 
thermal plasma component at $kT \la 60$ eV behind a line-of-sight 
column density. We suggest that these residuals are due 
to imperfectly-subtracted diffuse background emission, 
and are unrelated to the point-like emission from GRS\,1758$-$258, 
but we do not have enough data to prove it, also taking into account 
the additional uncertainties due to pile-up.

The RGS spectra (Figure 6) are consistent with the interpretation 
resulting from the EPIC spectra.
In 2001, the RGS spectrum is dominated by a thermal component 
with a small power-law excess at high energies. 
The RGS energy band is not sufficient to constrain 
the power-law component, so we fixed its slope and 
normalization to the best-fitting values derived from the MOS1 data 
(Table 5). We obtain that the best-fitting inner-disk temperature, 
normalization and total column density are also the same as in the 
MOS1 spectrum (Table 7). The small difference between our results 
and those of \citet{mil02} is due to our different choice 
of the power-law component.
In 2002, there is no need for a thermal component: 
the RGS spectrum is well fitted with an absorbed power law (Table 7). 
The photon index of the RGS spectrum is slightly steeper  
than in the simultaneous EPIC spectra, but marginally consistent 
considering the large error. Hints of a small soft excess 
below $0.9$ keV (both in 2001 and 2002) 
are consistent with what is seen in the EPIC-pn spectra.
We do not detect any significant features emission lines 
either in 2001 or in 2002.

\begin{table}
\begin{center}
\begin{tabular}{lrr}
\hline
Parameter & {2001 Value} & {2002 Value} \\
\hline\\[-5pt]
$N_{H,\rm Gal}$ & $0.75$ &  $0.75$ \\[5pt]
$N_{H,\rm int}$ & $1.00^{+0.04}_{-0.04}$ & $1.04^{+0.18}_{-0.17}$\\[5pt]
$kT_{\rm dbb}$ 
        & $0.45^{+0.02}_{-0.02}$ & --  \\[5pt]
$N_{\rm dbb}$ 
        & $1750^{+552}_{-417}$ & -- \\[5pt]
$\Gamma$ & $\left[2.85\right]$ & $2.02^{+0.47}_{-0.44}$\\[5pt]
$N_{\rm po}$ & $\left[0.054\right]$ &  $1.7^{+0.7}_{-0.5}$\\[5pt]
\hline\\[-5pt]
$\chi^2_{\nu}$ & $1.24~(580.0/469)$   & $1.02~(477.7/469)$ \\[5pt]
\hline 
\end{tabular} 
\end{center}
\caption{Best-fitting spectral parameters for the 2001 and 2002  
{\it XMM-Newton}/RGS observations. For the 2001 spectrum, the slope 
and normalization of the power-law component 
were fixed at the values found from MOS2. 
Units are as in Table 3.
Errors indicate the 90\% confidence interval for each parameter of interest.}
\label{tab7}
\end{table}

\subsection{X-ray time variability in 2000--2002}

We examined the variability of GRS 1758$-$258 in the three epochs.
For each epoch, we used MOS1 (in timing mode for the 2000 and 2002 
observations) and pn (which provides the highest count rate).
We extracted lightcurves binned to 0.01 s, 
from the same regions used for the spectral analysis.
We computed the Fourier transforms ({\it powspec} task in {\footnotesize FTOOLS}), 
normalized the resulting power spectra to be the fractional mean squared 
variability per unit frequency interval, and subtracted the Poisson noise level.

In the 2000 dataset, we found significant short-term variability 
below 10 Hz, with a power-law red noise component becoming 
dominant below $\sim 10^{-3}$ Hz, a flat top between $\sim 10^{-3}$--$1$ Hz, 
and a steepening above $\sim 1$ Hz. We do not have enough signal to noise
to determine whether there are quasi-periodic oscillations 
at $\sim 1$--$5$ Hz, as found by \citet{lin00} from the 1997 {\it RXTE} 
observations, but the MOS1 power spectrum hints at their presence 
(Figure 7, top panel).
The rms fractional variability between $10^{-4}$ and $50$ Hz 
is $(10 \pm 1)\%$. 
In 2001, the short term variability was much reduced: the 
rms fractional variability is consistent with zero. 
There may be a weak power-law noise component similar to the one seen 
in the 2000 lightcurves.
The 2002 observations show the highest degree of short-term variability, 
with rms $= (28 \pm 1)\%$. The strong band-limited noise 
has a break at $\approx 0.4$ Hz.
In summary, the timing properties of GRS 1758$-$258 in the three epochs 
are consistent with those of the canonical intermediate state (probably 
soft intermediate), high/soft state and low/hard state respectively 
\citep{bel96,men98,rei03,mot09,bel10}, 
in perfect agreement with the spectral identification.


\subsection{X-ray spectral state in 2008--2009}

The {\it Swift}/BAT hard X-ray monitoring during 2008--2009 
(Figure 2) suggests that the source was in the hard state 
during that time, before and after our ATCA observation.
The discontinuity in the X-ray monitoring between December 
and January is due to the unfavourable solar orientation, 
and does not represent significant evidence of state transitions.
The X-ray hard state is consistent with our radio core detection 
between the two epochs (Section 4.2).

The coadded 2008 April spectra have an on-source time 
of 21.1 ks in PC mode and 1.7 ks in WT mode. 
Both spectra are well fitted above 0.9 keV by a simple power-law 
model with photon index $\Gamma \approx 1.5$ (Table 8, Figure 8) 
and total column density $\approx 1.5 \times 10^{22}$ cm$^{-2}$. 
In fact, a simple power law is a good fit for all 
the (background-subtracted) PC-mode data, down to 0.5 keV. 
In WT mode, there is a small excess at energies $\la 0.9$ keV, similar 
to the excess found in some {\it XMM-Newton} spectra. 
We attribute this excess either to unsubtracted particle background 
or to extended thermal-plasma emission around the source.
Overall, the spectral parameters are very similar to those  
found in 2002 (Table 6 and section 6.1), but the luminosity is a factor of 2 lower.

The 2009 March spectra were taken only in WT mode.
Considering the systematic uncertainty of using a timing mode 
for spectral analysis, we found that both a simple power-law and 
a broken power-law model provide acceptable fits to the data 
above 0.9 keV (Table 9, Figure 8); the broken power-law is preferred, 
suggesting mild spectral curvature. For this reason, equally good fits 
can be obtained either with phenomenological combinations 
of a straight power-law and broad-band curved components 
in the {\sc XSPEC} arsenal, or with (more physical) 
Comptonization models \citep{loh11}. 
In any case, the photon index $\Gamma < 2$ over the whole fitting band 
is again typical of the hard state. 
There is, again, a small excess 
at energies $\la 0.9$ keV. That soft component is not consistent 
with a disk-blackbody spectrum behind the same column density 
as the power-law emission. As noted before (Section 6.1), 
it can formally be fitted with an optically-thin thermal plasma 
component at $kT \la 60$ eV and only line-of-sight column density. 
We do not think that this soft excess comes directly from GRS 1758$-$258.

In both the 2008 and 2009 {\it Swift} datasets, the total X-ray luminosity 
was $\approx 10^{37}$ erg s$^{-1}$, a factor of 2 lower than in 
the 2002 hard state observation. Therefore, in this case, the hard state 
luminosity was slightly lower than the 2001 soft state luminosity, 
unlike what is seen in other (more luminous) hard state observations of this BH.
This confirms that the spectral state of this BH is not solely 
determined by the total luminosity or mass accretion rate.
Based on these data alone, we cannot speculate about the physical structure 
of the accretion flow in the 2008--2009 hard state: {\it e.g.}, whether 
it was a truncated disk plus advective region, or a full disk 
covered by a hot corona or outflow. This issue will be 
investigated elsewhere \citep{loh11}.

\begin{figure}
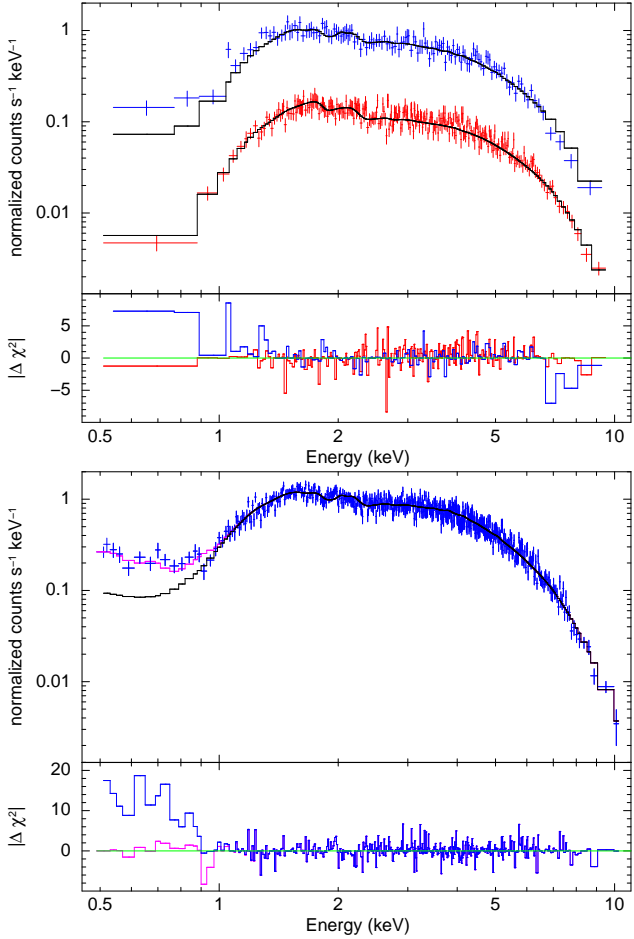

\begin{center}
\psfig{figure=2008_xrt.ps,width=62mm,angle=270}\\
\psfig{figure=2009_xrt.ps,width=62mm,angle=270}
\end{center}
\caption{Top panel: coadded {\it Swift}/XRT spectra from the four 
observations between 2008 April 01--11.
The spectrum in PC mode is plotted in red in the online version; the WT mode in blue.
The model is an absorbed power-law, fitted between $0.9$--$10$ keV 
for both spectra. See Table 8 for the best fitting parameters, 
and Section 6.3 for an explanation of the soft excess in the WT spectrum. 
Bottom panel: coadded {\it Swift}/XRT spectrum (WT mode) 
from the two observations of 2009 March 12--13. 
The model is a broken power law, fitted between $0.9$--$10$ keV.
See Table 9 for the best fitting parameters.
The soft excess is formally well fitted by a low-temperature 
thermal plasma component with only line-of-sight absorption (plotted 
in pink in the online version); see Section 6.3. }
\label{f8}
\end{figure}

\begin{table}
\begin{center}
\begin{tabular}{lrr}
\hline
Parameter & {PC Value} & {WT Value} \\
\hline
\multicolumn{3}{c}{Model: {\tt const*phabs*phabs*pow}} \\
\hline\\[-5pt]
Const & $1.00$ & $1.15^{+0.04}_{-0.03}$\\[5pt]
$N_{H,\rm Gal}$ & \multicolumn{2}{c}{$0.75$}  \\[5pt]
$N_{H,\rm int}$ & \multicolumn{2}{c}{$0.68^{+0.07}_{-0.07}$}\\[5pt]
$\Gamma$ & \multicolumn{2}{c}{$1.50^{+0.05}_{-0.05}$}\\[5pt]
$N_{\rm po}$ & \multicolumn{2}{c}{$0.49^{+0.04}_{-0.04}$}\\[5pt]
\hline\\[-5pt]
$f_{0.3-12}$ & $3.1^{+0.1}_{-0.1}$ 
        & $3.6^{+0.1}_{-0.1}$\\[5pt]
$L_{0.3-12}$ & $0.35^{+0.01}_{-0.01}$ 
        & $0.41^{+0.01}_{-0.01}$\\[5pt]
$L_{0.3-100}$ & $1.1^{+0.1}_{-0.1}$ & $1.3^{+0.1}_{-0.1}$\\[5pt]
\hline\\[-5pt]
$\chi^2_{\nu}$ &  \multicolumn{2}{c}{$0.90~(614.1/683)$}\\[5pt]
\hline
\end{tabular} 
\end{center}
\caption{Best-fitting spectral parameters for the coadded 2008 April
{\it Swift}/XRT observations. We allowed for a free normalization parameter 
between the PC and WT spectra. Units are as in Table 3.
Errors indicate the 90\% confidence interval for each parameter of interest.}
\label{tab8}
\end{table}

\begin{table}
\begin{center}
\begin{tabular}{lr}
\hline
Parameter & Value \\
\hline
\multicolumn{2}{c}{Model: {\tt phabs*phabs*pow}} \\
\hline\\[-5pt]
$N_{H,\rm Gal}$ & $0.75$  \\[5pt]
$N_{H,\rm int}$ & $0.71^{+0.06}_{-0.06}$\\[5pt]
$\Gamma$
        & $1.59^{+0.04}_{-0.04}$ \\[5pt]
$N_{\rm po}$ & $0.69^{+0.04}_{-0.04}$\\[5pt]
\hline\\[-5pt]
$f_{0.3-12}$ & $3.7^{+0.1}_{-0.1}$\\[5pt]
$L_{0.3-12}$ & $0.45^{+0.01}_{-0.01}$\\[5pt]
$L_{0.3-100}$ & $1.2^{+0.1}_{-0.1}$\\[5pt]
\hline\\[-5pt]
$\chi^2_{\nu}$ &  $1.18~(619.3/527)$ \\[5pt]
\hline
\hline
\multicolumn{2}{c}{Model: {\tt phabs*phabs*bknpow}} \\
\hline\\[-5pt]
$N_{H,\rm Gal}$ & $0.75$  \\[5pt]
$N_{H,\rm int}$ & $0.36^{+0.10}_{-0.09}$\\[5pt]
$\Gamma_1$
        & $1.12^{+0.11}_{-0.18}$ \\[5pt]
$E_{\rm b}$ 
        & $3.8^{+0.5}_{-0.4}$\\[5pt]
$\Gamma_2$ & $1.88^{+0.08}_{-0.14}$\\[5pt]
$N_{\rm po}$ & $0.40^{+0.06}_{-0.07}$\\[5pt]
\hline\\[-5pt]
$f_{0.3-12}$ & $3.5^{+0.1}_{-0.1}$\\[5pt]
$L_{0.3-12}$ & $0.36^{+0.01}_{-0.01}$\\[5pt]
$L_{0.3-100}$ & $0.8^{+0.1}_{-0.1}$\\[5pt]
\hline\\[-5pt]
$\chi^2_{\nu}$ &  $1.02~(536.8/525)$ \\[5pt]
\hline 
\end{tabular} 
\end{center}
\caption{Best-fitting spectral parameters for the coadded 2009 March 
{\it Swift}/XRT observations.  Units are as in Table 3.
Errors indicate the 90\% confidence interval for each parameter of interest. }
\label{tab9}
\end{table}

\section{Discussion}

\subsection{Disk size and BH mass}
The peak colour temperature of the disk is similar 
($kT_{\rm in} \approx 0.45$--$0.48$ keV) in the two epochs 
of our {\it XMM-Newton} observations 
when a disk component is directly visible (Tables 3,4,5).
A colour temperature $0.48$ keV was also found 
by \citet{pot06} from {\it INTEGRAL} observations 
in 2003 March--April, another epoch when the X-ray 
spectrum was dominated by the soft thermal component.
{\it RXTE}/PCA observations from 2001 February--March 
(a few days before our 2001 observations) also showed 
a dominant disk component with $kT_{\rm in} \approx 0.46$ keV 
\citep{smi01a}. A {\it Chandra}/ACIS HETG spectrum from 
2001 March 24 showed a peak colour temperature $\approx 0.51$ 
keV \citep{hei02b}.

The presence of a soft excess was suggested even when 
the source was in a harder, power-law dominated state.
{\it ROSAT}/PSPC observations from 1993 March 31--April 1 
are consistent with a thermal component with $kT_{\rm bb} \sim 0.5$ keV 
\citep{mer94,kec01}; 
in the {\it ASCA} observations of 1995 March 29--30, there are 
hints of a heavily Comptonized seed thermal component with 
$kT_{\rm bb} \sim 0.4$--$0.5$ keV \citep{mer97}. However, 
both observations can also be fitted with simple power laws 
\citep{mer97,kec01}.

The simplest explanation for this consistent value 
of the temperature is that it corresponds to the peak colour 
temperature of a full disk, extending all the way to the 
innermost stable circular orbit. If so, $T_{\rm in} \sim L^{1/4}$. 
Since the X-ray luminosity and probably the total accretion rate 
have varied only by a factor $\la 3$ over the last two decades, 
the peak temperature of a full disk (not significantly hardened 
by Comptonization) should vary only by a factor 
$\la 3^{0.25}$, that is only $\la 30\%$.
Most Galactic BH transients have typical disk temperatures 
$\approx 1$ keV in their high/soft state \citep{rem06}, 
with X-ray luminosities $\sim$ a few $10^{38}$ erg s$^{-1}$. 
GRS 1758$-$258 has an X-ray luminosity $\approx 2 \times 10^{37}$ 
erg s$^{-1}$; if its BH has a similar mass to the other 
Galactic BHs, we do indeed expect a disk temperature $\approx 0.5$ keV 
when all or most of the accretion power is directly 
radiated by the disk. That is the only state when 
we can constrain the BH mass from the disk temperature. 

We can use the disk normalization parameter $N_{\rm dbb}$ 
for a more quantitative estimate of its BH mass. The best-fit 
value $N_{\rm dbb} \approx 700$ found in the 2000 {\it XMM-Newton} dataset 
certainly underestimates the thermal emission from the full disk, 
because at least half of the disk photons have been 
upscattered into the power-law. This would not change the observed 
disk temperature but would reduce its flux normalization. 
In 2001, the disk component dominates and our most reliable spectral 
fits suggest $N_{\rm dbb} \approx 1700$. The 2003 {\it INTEGRAL} 
study gives $N_{\rm dbb} \approx 2700$ during another soft state \citep{pot06}. 
Let us take an average value of $2200$ for the sake of our ballpark 
mass estimate: then, the apparent inner radius $r_{\rm in}
\equiv d_{10} (N_{\rm dbb}/\cos \theta)^{0.5} \approx (40/\cos \theta)$ km.
The physical inner radius $R_{\rm in} \approx 
1.19 r_{\rm in} \approx (45/\cos \theta)$ km \citep{kub98}.
The inclinaton angle $\theta$ is unknown, but the source 
is unlikely to be perfectly face-on (given the extended jet lobe structure), 
or edge-on (because no X-ray eclipses have been seen). 
Assuming a Schwarzschild BH ($R_{\rm in} \approx 6M$) 
we obtain a characteristic mass $\approx \left(5/\cos \theta\right) M_{\odot}$, 
typical of Galactic stellar-mass BHs, and therefore an average 
X-ray luminosity $\sim 2\%$ Eddington. 

\subsection{Disk and corona during state transitions}

Taken individually, all three main spectral states 
of GRS 1758$-$258 are similar to the canonical states 
of Galactic BHs \citep{rem06,fen04}. The characteristic 
long-term X-ray luminosity $\approx 2 \times 10^{37}$ erg s$^{-1}$ 
is also consistent with the characteristic threshold 
for state transitions. However, the transition between 
states in GRS 1758$-$258 is unusual, as previously noted \citep{smi01a,pot06}. 
The radiatively-efficient soft state has a slightly lower total 
luminosity\footnote{The dramatic drop in the $2$--$10$ keV flux detected 
by {\it RXTE} (Figure 1) is largely compensated by the increase in flux 
from the disk, below $2$ keV.},  
and hence lower mass accretion rate, than during most (not all) of 
the long-duration hard state.
For example, our {\it XMM-Newton} spectral analysis shows that 
during the 2001 soft state, the source was slightly less luminous 
than in the harder states of 2000 and 2002, although slightly 
more luminous than in the 2008--2009 hard state. 
Moreover, the disk component has been seen to appear and disappear 
over very short timescales (Table 10), but always with a peak colour 
temperature $\approx 0.45$--$0.50$ keV.
This is difficult to reconcile with the disk truncation/refill scenario 
\citep{esi97,esi98} that is often taken as the standard paradigm 
for hard/soft transitions in Galactic BHs.

\begin{table*}
\begin{center}
\begin{tabular}{lccccr}
\hline
Epoch & Spectral State & Disk Temperature & Power-law Slope & Radio Core? & References\\
\hline\\[-5pt]
2000 Jan--Aug & Hard & disc not seen & $\approx 1.5$--$1.8$ &  ? & [1] \\
2000 Sep & Int & $\approx 0.5$ keV & $\approx 2.0$--$2.6$ & ?  & [1,2] \\
2000 Oct--Dec & Hard & disc not seen  & $\approx 1.5$--$1.8$ &  ? & [1] \\
2001 Jan 23 & Int & $\approx 0.5$ keV & $\approx 2.2$ &  ? & [1,3] \\
2001 Jan 29--Feb 21 & Hard & disc not seen  & $\approx 1.8$--$2.0$ &  NO & [1,3] \\
2001 Feb 27 & Soft & $\approx 0.5$ keV & $\approx 2.6$ &  ? & [3,4] \\
2001 Mar 2--24 & Soft & $\approx 0.5$ keV & $\approx 2.8$--$3.2$ &  NO & [2,3,4,5] \\
2001 Apr--Nov & Soft & ? & ? & NO & [2,6,7] \\
2002 Jan--Mar & Int? & ? & ? & NO & [2,4,7] \\
2002 Jul--2003 Feb & Hard & ? & ? & YES & [2,6,7] \\
2002 Sep 28  & Hard & disc not seen  & $\approx 1.5$--$1.7$ & YES? & [2] \\
2003 Mar--Apr & Soft/Int & $\approx 0.5$ keV & $\approx 2.3$ & ?  & [6,8] \\[5pt]
\hline 
\end{tabular} 
\end{center}
\caption{Timeline of the 2000--early 2003 state transitions. ``Disc not seen'' means 
that a thermal disk component is not required to fit the X-ray spectrum 
at those epochs. References:
[1] = \citet{smi02a}, [2] = this work, [3] = \citet{smi01a}, 
[4] = \citet{smi01b}, [5] = \citet{hei02b}, [6] =\citet{pot06}, [7] = \citet{hei02a}, 
[8] = \citet{pot08}.}
\label{tab10}
\end{table*}

As an alternative explanation, \citet{smi01a,smi02a} suggested 
that a full accretion disk is already present before the hard-to-soft 
transition, and is still present at least for some time after 
the return to the hard state. In this scenario, sometimes the disk is directly visible, 
but more often it is completely covered by a Comptonizing medium above it. 
The presence of full disks with inner radii close to the innermost stable 
circular orbit even in the low/hard state, at luminosities $\sim 0.01$ Eddington, 
was suggested by \citet{rei10} from their study of several Galactic BH 
transients. Their findings suggest that transitions from the high/soft state 
to the low/hard state may be driven by changes in the corona 
(perhaps related to jet formation) rather than by the disappearance 
of the accretion disc. The behaviour of GRS 1758$-$258 suggests 
the same scenario for transitions in the both directions, 
from hard to soft and vice versa, over short periods of time. 

One mechanism to produce a rapidly changing corona over a steady disk 
is the two-flow scenario \citep{smi01a,smi02a,cha95}: a keplerian accretion flow 
feeds the disk (evolving over a viscous timescale) 
and an independent sub-keplerian flow feeds the hot Compton cloud 
(evolving over a dynamical timescale). Soft states correspond 
to the temporary shutoff of the sub-keplerian flow. 
One difficulty of this scenario is that the donor is 
a low-mass star, an unlikely source for a low-angular-momentum  
component of the accretion flow.

We suggest another possibility: the hard emission component 
may come from a magnetically-powered coronal outflow. Radio studies provide 
independent evidence that there is a jet during the hard states. 
Coronal magnetic fields may dissipate a large fraction of accretion power 
and transport it out as Poynting flux or as a magnetically driven wind 
\citep{mer02,kun04}. In this scenario, the accretion flow 
may switch between a standard disk and a coronal outflow 
in response to changes in the configuration of the poloidal magnetic field, 
and hence in the fraction of angular momentum transported 
by the azimuthal-vertical component of the magnetic stress 
rather than by the radial component. The underlying full disk 
may always be present even during the outflow dominated 
phases, but it will be masked by the outflow, 
and also colder and less luminous than in 
the disk dominated phases, for the same mass inflow rate \citep{jol08}. 
Switches between a colder and a hotter disk phase may occur 
on the thermal timescale, which is shorter than 
the viscous timescale required to regrow a disk from large radii
\citep[$t_{\rm th} \sim (H/R)^2 t_{\rm visc} \ll t_{\rm visc}$:][]{fkr02}.
Magnetohydrodynamic simulations suggest 
\citep{dev03,dev05,haw06} that magnetized outflows are not isotropic:  
they are confined by funnel walls. In this scenario, we speculate that 
for GRS 1758$-$258 we are looking down the funnel, so that 
the apparent luminosity of the hard state is enhanced by 
a factor of a few compared with the luminosity in the soft 
thermal state (which has a more isotropic emission).
This would make the true luminosity and accretion rate 
of the hard state lower than in the soft state, 
as observed in most other Galactic BHs.


\subsection{Radio louder and radio quieter microquasars}

There is now strong evidence, from both the 1992 and 2001 
state transitions, that the jet in GRS 1758$-$258 
switches off during soft states, similarly to most other 
stellar-mass BH transients. This is the basis of the canonical 
model for the disc-jet coupling in black hole X-ray binaries 
\citep{fen04} and AGN \citep{kor06}. 
When accreting BHs are in the jet-dominated state, 
there is a fundamental correlation between their X-ray luminosity 
and radio core luminosity \citep{mer03,gal03,fal04}.
As more and more Galactic BHs have been studied in radio 
and X-ray bands, it was found \citep{gal07,sol10,cal10,sol11} 
that there are two distinct groups or sequences of sources, 
with similar X-ray luminosity and hard spectral state 
but radio luminosities differing by a factor $\sim 10$--$30$.

GRS 1758$-$258 belongs to the radio-quieter group, despite 
its characteristic large-scale radio jet and lobes. In fact, 
this source is the radio quietest of all Galactic BHs 
in that X-ray luminosity range, $\approx 50$--$100$ times fainter 
that the radio-loud sequence (Figure 9). Even if its X-ray luminosity 
were revised downwards by a factor of a few due to geometrical 
beaming (Section 7.2), GRS 1758$-$258 would still be in 
the radio-quieter group.

The existence of radio-loud and radio-quiet stellar-mass BHs 
is reminiscent of the AGN classification. For a given range 
of black hole masses and accretion rates, some kinds of active galaxies 
have stronger jets than others \citep{wil95,sik07}. In the case of AGN, 
this dichotomy corresponds to morphological differences in the host 
galaxy: elliptical galaxies are generally more radio-loud than disk galaxies. 
It is still not clear what corresponding environmental 
parameter is associated to the radio dichotomy in stellar-mass BHs: 
no obvious dependence was found on orbital periods, disc sizes, 
outburst history \citep{sol11}.

Spin is often suggested as the intrinsic physical difference 
between radio loud and radio quiet BHs: fast-spinning black 
holes can launch more powerful jets \citep{bla90}.
This is consistent with the scenario that black holes in giant 
elliptical galaxies evolved to higher spins than those in spiral galaxies.
However, we do not have enough reliable spin measurements 
for stellar-mass BHs to test whether the same correlation applies there.
It is also still unclear whether the jet power has a direct contribution 
from the BH spin \citep{bla77,mcn10,fen10}. Furthermore, \citet{bro11} 
suggest that BH spin has considerably less impact on AGN jet power 
than has been previously reported.

Alternatively, the radio dichotomy in stellar-mass BHs 
may reflect different modes of disc/jet coupling \citep{sol11}, 
or different structures of the accretion flow in the hard state.  
For example, the efficiency of jet launching  
may be different between an ADAF-like configuration (truncated disk 
and hot, advective inner region) and a disc-plus-corona structure 
(corona on top of a full disk). Both accretion states 
may be phenomenologically classified as ``low/hard state'' 
with similar X-ray properties, but different radio properties.

We note that the observed presence of two distinct sequences 
of Galactic BHs in the radio/X-ray luminosity plane (Figure 9) 
is similar to the dichotomy between radio galaxies (including 
both FRII and FRI objects) and Seyfert galaxies. Both radio 
galaxies and Seyferts show a strong correlation between core radio 
and X-ray luminosity (Fig.~2 in \citealt{pan07}), 
or core radio and [OIII] luminosity (Fig.~4 in \citealt{bic02}), 
which is another proxy for the radiative accretion luminosity.
But the two sequences are shifted by almost three orders 
of magnitude, with Seyfert galaxies being fainter 
in the radio for a given X-ray luminosity. Even after both samples 
are normalized by their masses, according to the fundamental 
plane relations \citep{mer03}, they still form two distinct 
sequences, shifted by a factor $\sim 50$ in radio luminosity 
(Fig.~7, right panel, in \citealt{pan07}).
Similarly, \citet{bro11} plotted radio loudness (also determined 
using core radio luminosities, and with a mass correction applied) 
as a function of the Eddington ratio. They found that radio galaxies 
are on average more radio-loud than Seyfert galaxies and LINERS 
by a factor of $\sim 30$ for Eddington ratios $< 1$ per cent.

By analogy with the model of \citet{bic02}, we suggest 
that this dichotomy corresponds to the different 
radio efficiency of relativistic electron-positron jets 
(the radio loud sequence) and sub-relativistic, thermally 
dominated baryonic jets (the radio quiet sequence).
In this scenario, there is still a universal correlation 
between jet power and radiative luminosity, but the same 
jet power corresponds to different radio outputs 
in fast leptonic and slow mass-loaded jets.
Possible observational tests of this hypothesis  
include measurements of circular polarization 
in the radio core, and of the location and intensity 
of the 511 keV annihilation line. For sources where the jet 
creates a plasma cocoon around the BH, comparison 
of the radio luminosity with the mechanical power necessary 
to inflate the bubble can also constrain the fraction 
of jet energy carried by protons and ions.

\begin{figure}
\begin{center}
\psfig{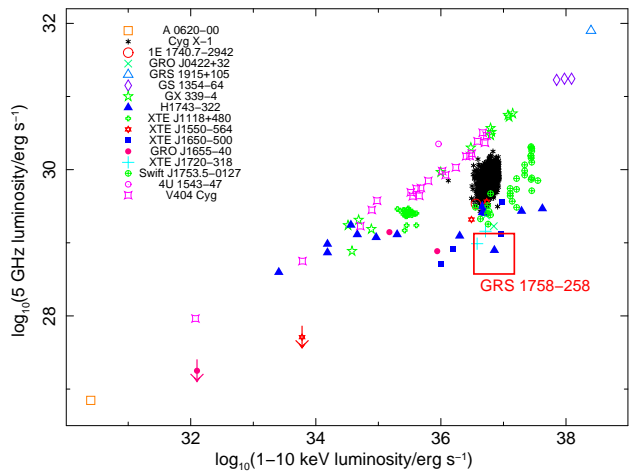}
\end{center}
\caption{Radio loud and radio quiet sequences of Galactic BHs.
The red box marks the characteristic range of X-ray and radio variability 
of GRS 1758$-$258 during the hard state (when the radio core is detected). 
Datapoints from all other sources are from \citet{cal10}.}
\label{f9}
\end{figure}

\subsection{Extended radio lobes}

Our 2008--2009 ATCA observations confirm the presence of extended 
radio lobes at apparently the same position as in 1992 and 1997, 
within the uncertainty due to different beam sizes and sensitivities.
Such structures are common in radio galaxies  
but much rarer and fainter in microquasars. 
It was suggested \citep{heinz02} that microquasars lack persistent lobes 
because they are mostly located in a low-density environment 
compared with AGN (scaled to their respective jet powers), 
so that any lobes formed have short lifetimes against adiabatic losses.
Ejections of fast-moving synchrotron-emitting blobs are well documented 
in transient Galactic BHs such as GRS 1915$+$105 \citep{rod99}, 
GRO J1655$-$40 \citep{tin95}, and XTE J1550$-$564 \citep{cor02}; 
see also \citet{mir99} for a comprehensive review.  
Such ejecta tend to decelerate and fade over shorter timescales 
(months or few years), after travelling a characteristic 
distance $\approx 0.5$ pc \citep{hao09}.

The distance between the core and the lobes of GRS 1758$-$258 
is an order of magnitude larger than that. Thus, they are unlikely 
to be individual blob ejections from a historical (unrecorded) outburst, 
unless the source is located in a very large cavity, that allowed 
the ejections to move that far before stopping. 
A more likely explanation is that the lobes correspond to the locations 
where the heads of a pair of continuous jets (moving in opposite directions) 
encounter the interstellar medium and dissipate 
their bulk kinetic energy. Our new radio observations 
confirm the energy content inferred by \citet{har05}, 
that is $\sim$ a few $10^{45}$ erg in relativistic electrons.
If the jet power is similar to the radiative core power, 
that is $P_{\rm jet} \sim P_{\rm rad} \sim 10^{37}$ erg s$^{-1}$ 
(as is the case for example in Cyg X-1: \citealt{rus07}), such energy  
can be supplied in a few years. This would also imply that 
the lobe energy content, and therefore radio flux, can vary  
on a characteristic timescale of a few years, in response 
to changes in the injection power. 

However, this estimate is very uncertain because we do not yet know 
the fraction of kinetic and thermal energy stored in protons, ions 
and non-relativistic electrons. In several extra-Galactic jet sources, 
an independent estimate of the integrated jet power is available from 
the size of the jet-inflated cocoon around the BH. In those cases, 
the total jet power can be up to a few hundred times higher 
than the power transferred to the radio-emitting 
relativistic electrons \citep{cav10,sor10,pak10}.
Moreover, the radio core of GRS 1758$-$258 is $50$ to $100$ times 
fainter, at a given X-ray luminosity,  
than the ``standard'' radio-loud class of Galactic BHs 
\citep[Section 7.3;][]{sol10,sol11}. 
We do not know whether that means that the total kinetic jet power 
(including both leptons and baryons) is correspondingly lower, 
or only the fraction of kinetic power carried into the lobes 
by the relativistic electrons is lower.
But in either case, the kinetic power energizing the radio-emitting 
electrons in the lobes is more likely to be $\la 10^{35}$ erg s$^{-1}$.
corresponding to a timescale $\ga 10^3$ yrs, comparable to 
the flux decay timescale due to adiabatic expansion \citep{har05}.

\section{Conclusions}

We have studied the Galactic BH GRS 1758$-$258 in the X-ray and radio bands, 
during its spectral state transitions of 2000--2002, and during 
the hard state of 2008--2009. We showed that a compact jet is present 
during the hard state, and switches off in the soft state.
We identified the X-ray states from their spectral and timing properties, 
constrained the disk parameters during the soft and intermediate state, 
and estimated a BH mass $\sim 10 M_{\odot}$ and an X-ray luminosity 
$\approx 0.01$--$0.03 L_{\rm Edd}$.

We discussed its unusual X-ray state behaviour: the apparent luminosities 
in the hard and soft states overlap; in fact, it is often higher 
in the hard state. Since radiative efficiency in the hard state 
must be at least slightly lower than in the soft (thermal) state, 
this implies that the mass accretion rate in the hard state is often 
higher than in the soft state (assuming there is no beaming).
This suggests that another physical parameter 
is more important than changes in the total mass accretion rate, 
for driving state transitions. 
The sequence of spectral variability in 2001 also suggests 
that a full disk was already present at least for several weeks 
before the main soft transition, when the X-ray source was 
in a hard or intermediate state.
We conclude that the hard state of GRS 1758$-$258 may not be the same 
as the low/hard state of other BH transients successfully modelled 
with truncated disks plus ADAFs; it may be an example 
of a Comptonizing corona/outflow on top of the disk.

We showed that the radio lobes, first seen in 1992, were still 
present in 2009, apparently at the same location. This supports 
their interpretation as long-lived relic structures, perhaps 
associated with the interaction of the jet with the interstellar 
medium. We are leaving a more detailed study of the lobe evolution 
to a planned long-term ATCA radio monitoring of this source over 
the next few years. 
We measured the radio core luminosity in a series of unpublished VLA 
monitoring observations over 2001--2003, and compared it with the radio 
core fluxes measured in 1992--1993, 1997 and 2008--2009. When the compact jet 
is detected, the 5-GHz flux typically varies between $\approx 0.1$--$0.5$ mJy.
Comparing this with the X-ray luminosity, we showed that 
GRS 1758$-$258 belongs to the radio-faint sequence of Galactic BHs 
in the radio/X-ray plane. The dichotomy between radio-loud and radio-faint 
stellar-mass BHs for the same X-ray luminosity and spectral state 
is reminiscent of a similar dichotomy between Seyfert and radio galaxies 
in the fundamental plane. BH spin is often suggested as a possible 
explanation, with fast-spinning BHs able to launch more powerful jets 
at the same accretion rate. We suggested an alternative explanation: 
both the radio-loud and radio quiet sequences may have the same jet power 
at a given accretion rate, but radio-loud BHs have a relativistic 
electron/positron jet, and radio-quiet BHs (such as GRS 1758$-$258) 
a sub-relativistic, mass-loaded jet, which is less efficient 
at emitting synchrotron radiation for the same kinetic power.

\section*{Acknowledgments}

We thank Maria Diaz-Trigo, Rob Fender, 
James Miller-Jones, Manfred Pakull, David Smith, Kinwah Wu 
for useful discussions, and Tasso Tzioumis for assistance 
with the ATCA observations. RS is grateful for the financial support 
from Tsinghua University (Beijing) during part of this research.  
JWB acknowledges support from the UK Science and Technology Facilities Council.


\end{document}